\documentclass[graphicx,11pt]{aastex}

\lefthead{}
\righthead{}

\usepackage{pdfpages}
\begin{document}
%\linenumbers

\title{Subsolar Al/Si and Mg/Si ratios of non-carbonaceous chondrites reveal planetesimal formation during early condensation in the protoplanetary disk} 

\author{\textbf{A. Morbidelli$^{(1)}$, G. Libourel$^{(1)}$, H. Palme$^{(2)}$,  S.~A. Jacobson$^{(3)}$ and D.~C. Rubie$^{(4)}$}\\  
(1) Laboratoire Lagrange, UMR7293, Universit\'e de Nice Sophia-Antipolis,
  CNRS, Observatoire de la C\^ote d'Azur. Boulevard de l'Observatoire,
  06304 Nice Cedex 4, France. (Email: morby@oca.eu / Fax:
  +33-4-92003118) \\
  (2) Senckenberg, world of biodiversity, Sektion Meteoritenforschung. Senckenberganlage 25, 60325 Frankfurt am Main, Germany\\
  (3) Northwestern University, Dept. of Earth and Planetary Sciences, Evanston, 60208 Illinois\\
  (4) Bayerisches Geoinstitut, University of Bayreuth, 95440, Bayreuth, Germany\\
} 

\begin{abstract}
The Al/Si and Mg/Si ratios in non-carbonaceous chondrites are lower than the solar (i.e., CI-chondritic) values, {in sharp contrast to the non-CI carbonaceous meteorites and the Earth, which are enriched in refractory elements and have  Mg/Si ratios that are solar or larger}. We show that the formation of a first generation of planetesimals during the condensation of refractory elements implies the subsequent formation of residual condensates with strongly sub-solar Al/Si and Mg/Si ratios. The mixing of  residual condensates with different amounts of material with solar refractory/Si element ratios explains the Al/Si and Mg/Si values of non-carbonaceous chondrites. To match quantitatively the observed ratios, we find that the first-planetesimals should have accreted when the disk temperature was $\sim 1,330$--$1,400$~K {depending on pressure and assuming a solar C/O ratio of the disk}. We discuss how this model relates to our current understanding of disk evolution, grain dynamics, and planetesimal formation. We also extend the discussion to moderately volatile elements (e.g., Na), {explaining how it may be possible that the depletion of these elements in non-carbonaceous chondrites is correlated with the depletion of refractory elements (e.g., Al).} {Extending the analysis to Cr, we find evidence for a higher than solar C/O ratio in the protosolar disk's gas when/where condensation from a fractionated gas occurred.} {Finally, we discuss the possibility that the supra-solar Al/Si and Mg/Si ratios of the Earth are due to the accretion of $\sim 40$\% of the mass of our planet from the first-generation of  refractory-rich planetesimals}. 
\end{abstract}

\section{Introduction}

\begin{figure}[t!]
\centerline{\includegraphics[height=7.cm]{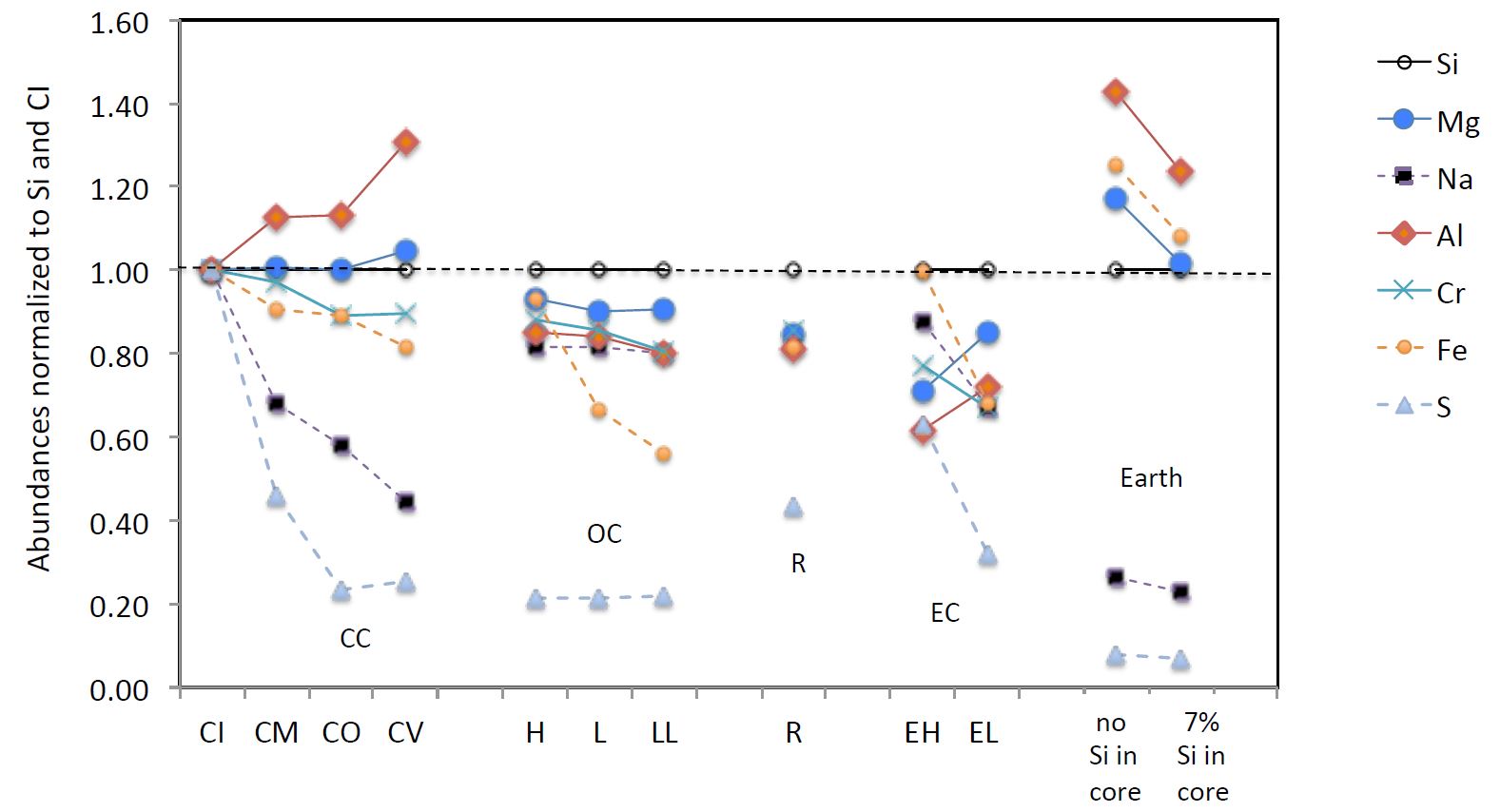}}
\caption{\linespread{0.5}\footnotesize Abundances of Al, Mg, Fe, Cr, Na, S, relative to Si and normalized to the CI ratios for different chondrites and the Earth (data and references reported in table~S1). For the bulk Earth (BE) two cases are given assuming 0 and 7wt\% of Si in the core respectively. There is a clear difference in chemistry between the carbonaceous chondrites (CC) and the non-carbonaceous chondrites.}
\label{Palme}
\end{figure}

{Carbonaceous chondrites (CCs), from CI to CM, CO and CV, show a progressive increase in Al/Si and Mg/Si ratios together with an increasing depletion in moderately volatile elements (e.g., Na) - see Fig.~\ref{Palme}, table~ S1 and references therein. CI meteorites reflect the solar composition and the observed trend is interpreted as the consequence of an increasing abundance of minerals formed in high-temperature regions of the protoplanetary disk (e.g., CAIs: Hezel et al., 2008). The trend continues for the Earth, {if the upper mantle Mg/Si and Al/Si ratios are valid for the bulk mantle, which is often assumed  (e.g., Palme and O'Neill, 2014; Hyung et al., 2016), but there are also arguments against a chemically uniform mantle (e.g.,  Ballmer et al., 2013). There is also some discussion} on the amount of Si in the core (Rubie et al., 2015). Thus, in Fig.~\ref{Palme} we report two values for the bulk Al/Si and Mg/Si ratios, computed assuming that the mantle is homogeneous (Hyung et al., 2016) but the core contains 0\% or 7\% Si by mass. This is intended to give a sense of the systematic uncertainty on the bulk composition ratios of the Earth. It remains clear, however, that the Earth is enriched in refractory elements and depleted in moderately volatile elements at least at the level of CO-CV meteorites. 

Enstatite chondrites, {rumuruti} and ordinary chondrites (ECs, RCs and OCs hereafter, {generically denoted NCCs for non-carbonaceous chondrites}) surprisingly do not follow this trend, despite the fact that they should have formed in the relatively hot, inner solar system. Their Al/Si and Mg/Si ratios are sub-solar while they have only a small Na depletion compared to COs, CVs and the Earth (Fig.~\ref{Palme}).}

The strong difference in Al/Si and Mg/Si ratios between the Earth and ECs poses a conundrum. {Most elements have the same  ratios of stable isotopes (e.g., O-isotopes) in ECs and Earth, whereas other chondritic meteorites show significant differences. Small nucleosynthetic anomalies for ECs relative to Earth have been reported for Mo, Ru and Nd.} For this reason  it is sometimes suggested that the Earth accreted mostly from enstatite chondrites ({see for instance} Javoy, 1995; {Lodders, 2000}, Dauphas, 2017). But the large differences in Al/Si and Mg/Si ratios between the Earth and ECs seem to preclude this possibility, {although some solutions have been proposed (e.g., incongruent vaporization of enstatite: see Mysen and Kushiro, 1988)}. Dauphas (2017) postulated that the precursors of the Earth  had the same isotopic composition as ECs but a different element composition. 

{It has been proposed that the sub-solar Al/Si and Mg/Si ratios in the NCCs are due to the loss of a refractory-rich component from a disk with an original solar composition (see Larimer, 1979; see also Alexander, 2019 and references therein). It has also been suggested that this refractory component was accreted by the Earth, thus explaining its enrichment in refractory elements. For instance, to explain the Mg-enrichment of the Earth, Dauphas et al. (2015) suggested that the ECs were depleted in a forsterite condensate and are complementary to the forsterite enriched Earth. This possibility has been explicitly excluded by Alexander (2019), for the reasons discussed in Sect.~\ref{Earth}.} An alternative explanation for the relative deficiency of Si and Mg in the Earth is the evaporation from the molten surface of the protoplanetary embryos that made our planet (Pringle et al., 2014; {Mysen and Kushiro, 1988}; Hin et al., 2017; Young et al., 2019). However, the sub-solar Al/Si and Mg/Si ratios of the NCCs cannot be explained by evaporation models because Si should be lost preferentially relative to Mg and Al. 

{In this paper, we revisit the idea of the sequestration of a refractory component from the protoplanetary disk in order to explain the depletion of refractory elements in the NCCs. In section~\ref{context}, we discuss  how a refractory-rich component could be sequestered, in the framework of modern models of disk evolution and planetesimal formation. In section~\ref{char}, we determine the temperature at which this refractory-rich component should have been isolated from the condensation sequence in order to explain the Al/Si and Mg/Si of the NCCs. {In section~\ref{others}, we extend our considerations to other elements: Fe, Na, Cr and S.} In section~\ref{Earth}, we show that the enrichment in refractory elements of the Earth can be explained, within  uncertainties, by the addition of the refractory component removed from the NCCs. We discuss how the issues that led Alexander (2019) to exclude this possibility can be solved, so that this possibility remains the most viable one. Section~\ref{Conclusions} summarizes the results and highlights the predictions of the properties of the protoplanetary disk that our results imply, that will need to be validated by future models of disk's formation and evolution.}    

\section{An astrophysical scenario for the sequestration of refractory material}
\label{context} 

As often assumed, we consider that the inner disk was initially very hot, so all material was in a gaseous form, including refractory elements. As the disk was rapidly cooling, various species started to condense in sequence, from the most refractory to the more volatile. This is typical of equilibrium condensation where, with declining temperature,  some earlier condensed species are destroyed to form new ones, in addition to condensing new species. 

{However, it is unlikely that this process continued undisturbed to low temperatures}.  When enough solid grains condensed, but well before the completion of the whole sequence, the ionization level of the gas was strongly reduced by the now present dust and the magneto-rotational instability (MRI) generating turbulence in the midplane of the disk was quenched (Desch and Turner, 2015). {Because the disk had a radial temperature gradient, this did not happen everywhere at the same time. The innermost part of the disk, hot enough to contain only small amounts of dust, remained MRI-active while the outer part located beyond a threshold distance corresponding to some temperature $T_{MRI}$ was MRI-inactive; the boundary between the two parts moved sunwards as the disk cooled with time. The value of $T_{MRI}$ is not known precisely. In this work, we will consider it as a free parameter {and we discuss its best-fit value in section~\ref{cs}.}}

The transition from a MRI-active to a MRI-inactive region of the disk should correspond to a transition in gas density. The surface density of gas has to be larger in the MRI-inactive region for the conservation of mass flux: $F_M=2\pi r \Sigma v_r$, where $F_M$ is the mass flux, $r$ is the distance from the star, $\Sigma$ is the surface density of gas and $v_r$ its radial velocity. In fact, $v_r=-3/2(\nu/r)$ where $\nu$ is the gas viscosity, so there is an anti-correlation between $\Sigma$ and $\nu$, the latter being much larger in the MRI-active region of the disk. This steep density gradient, creating a pressure maximum, acts as a barrier to the inward radial drift of dust due to gas drag. Thus, at the boundary between the MRI-active and MRI-inactive parts of the disk we expect to find not just the locally-condensed grains corresponding to that temperature, but also grains that condensed earlier and farther away and whose radial migration was stopped at this boundary ({Flock et al., 2017}). Incidentally, these migrated grains trapped at $T_{MRI}$ would evolve to acquire the same chemistry as those condensed locally at $T_{MRI}$. Hence, the dust/gas ratio at $T_{MRI}$ can become much larger with time than the value due solely to the locally-condensed grains.

This accumulation of dust  ({Flock et al., 2017}) should have triggered rapid planetesimal formation via the streaming instability (Youdin and Goodman, 2005; Johansen and Youdin, 2007). It is indeed known that this instability is triggered when the solid/gas mass ratio exceeds by 3-4 times the value corresponding to the condensation of dust from a gas of solar composition (the exact value depending on particle size; Yang et al., 2017). With the formation of planetesimals, the already condensed material stops participating in the gas-solid equilibrium chemistry because it becomes locked-up in large objects instead of small grains. 

Consequently, at the temperature $T_{MRI}$ there should have been a bifurcation in the chemistry of the disk due to the fractionation of solids from gas.  Below $T_{MRI}$, condensation could form new solid species only from the {\it residual} gas. Consequently, the first generation of planetesimals should have been refractory-rich, with strongly supra-solar Al/Si and Mg/Si ratios, whereas the second generation of solids condensed from the residual gas (denoted ``residual condensates'' hereafter for brevity) should have been extremely refractory-poor, with strongly sub-solar Al/Si and Mg/Si ratios. 

\begin{figure}[t!]
\centerline{\includegraphics[height=11.cm]{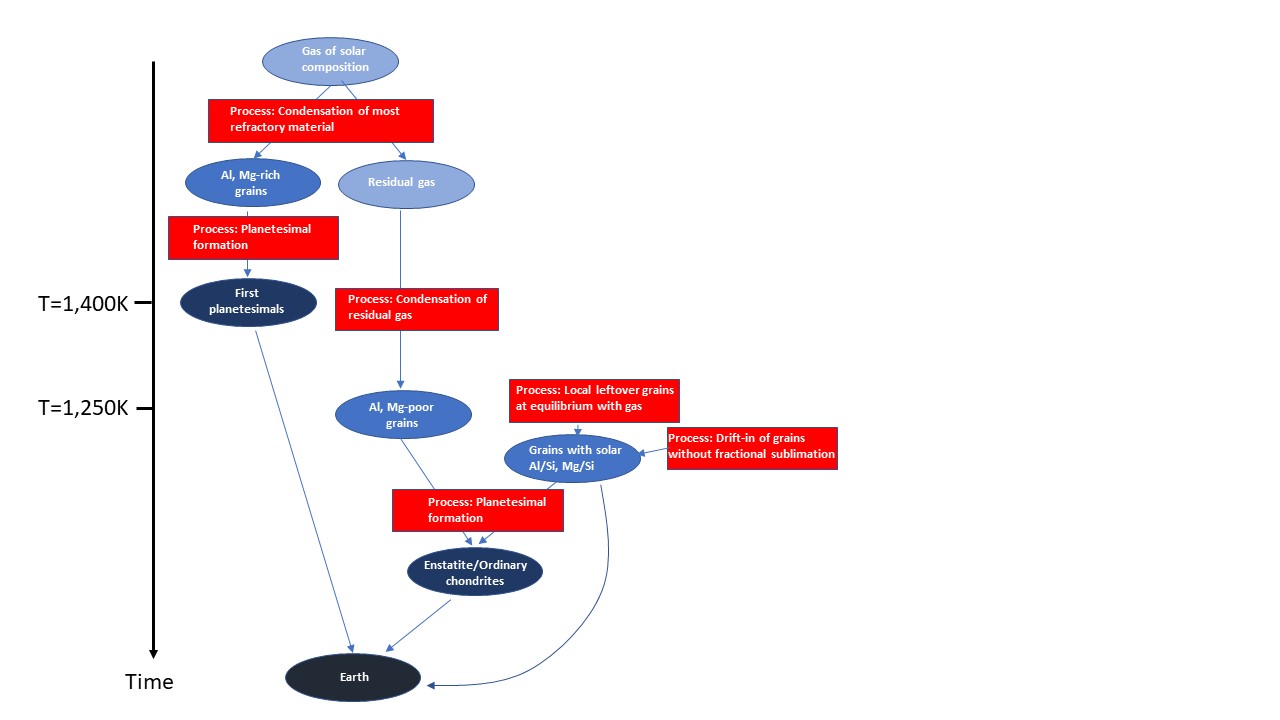}}
\caption{\linespread{0.5}\footnotesize A flow-chart of the scenario explored in this paper. The vertical axis represents the evolution of time, from top to bottom, during which the temperature decreases. The blue ellipses show the material  available (light blue for gas then -with increasing intensity of the color- grains, planetesimals and the Earth) and the red rectangles show the transformation processes. For simplicity, we do not track volatiles condensing below 1,250K, nor H and He. See  also Supplementary Fig. 3 for a more comprehensive (but complicated!) sketch of the evolution of gas and solids in the disk as a function of distance, time and declining temperature.}
\label{sketch}
\end{figure}

{This sequence of events and the roles of residual condensates, first planetesimals and solar-composition material in making the NCCs and the Earth are depicted in Fig.~\ref{sketch} and are described in more details in the subsequent sections.} 

\section{Determining the temperature of sequestration of refractory material}
\label{char}

{To determine $T_{MRI}$, the temperature of sequestration of the first condensates in planetesimals, we follow the condensation sequence from a gas of solar composition to track the species condensed and the chemical composition of the residual gas as a function of temperature (sect.~\ref{cs}). Then, we use these data to determine the temperature that fits best the observed Al/Si and Mg/Si ratios of the NCCs (sect.~\ref{results}).} 

\subsection{Condensation sequence}
\label{cs}

We have computed a condensation sequence, during the cooling of the disk at the mid-plane of the Solar Nebula, starting from 2,000~K and assuming a total pressure of $10^{-3}$ bar {and a solar C/O ratio}. {Some results for different pressures and C/O ratios are also reported below.} These calculations describe the equilibrium distribution of the elements and their compounds between coexisting phases (solids, liquid, vapor) in a closed chemical system with vapor always present upon cooling. We have used for this proof of concept a simplified solar gas composition taken from Ebel and Grossman (2000) composed of the most abundant elements of the solar photosphere: H, C, O, Mg, Si, Fe, Ni, Al, Ca, { Ti, Cr, S} and Na.  

The equilibrium condensation sequence of this simplified solar gas has been calculated using the FactSage software package (Bale et al., 2016) by means of a general Gibbs free energy minimization method (Eriksson and Hack, 1990). The thermodynamic data for each compound are taken from the database provided by the FactSage package. {The potential number of species that can be formed from the combination of the 13 elements considered in this system is over 450, comprising 150 gases and 320 pure solid phases (see tables~S2,~S3). For this proof of concept, condensation calculations have been performed in steps of 10~K, allowing only pure solid phases to condense, no solid solutions (see below). } 

\begin{figure}[t!]
  \centerline{\includegraphics[height=6.cm]{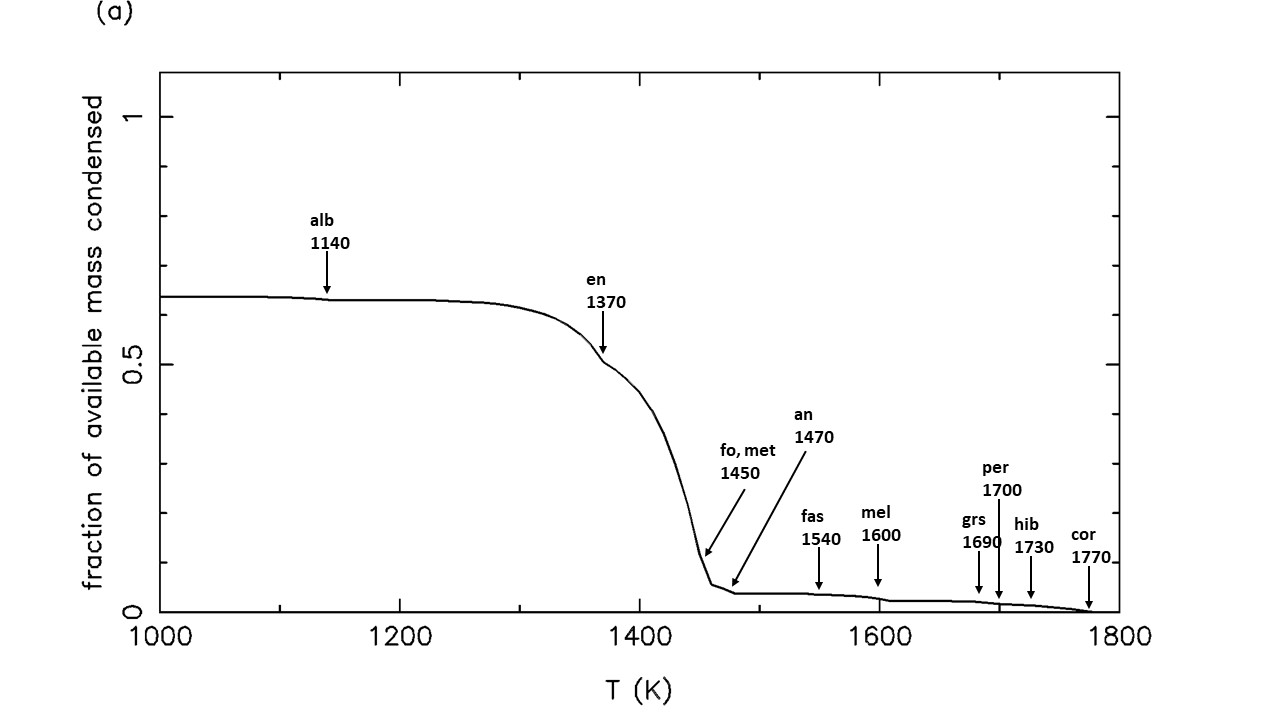}}
  \centerline{\includegraphics[height=6.cm]{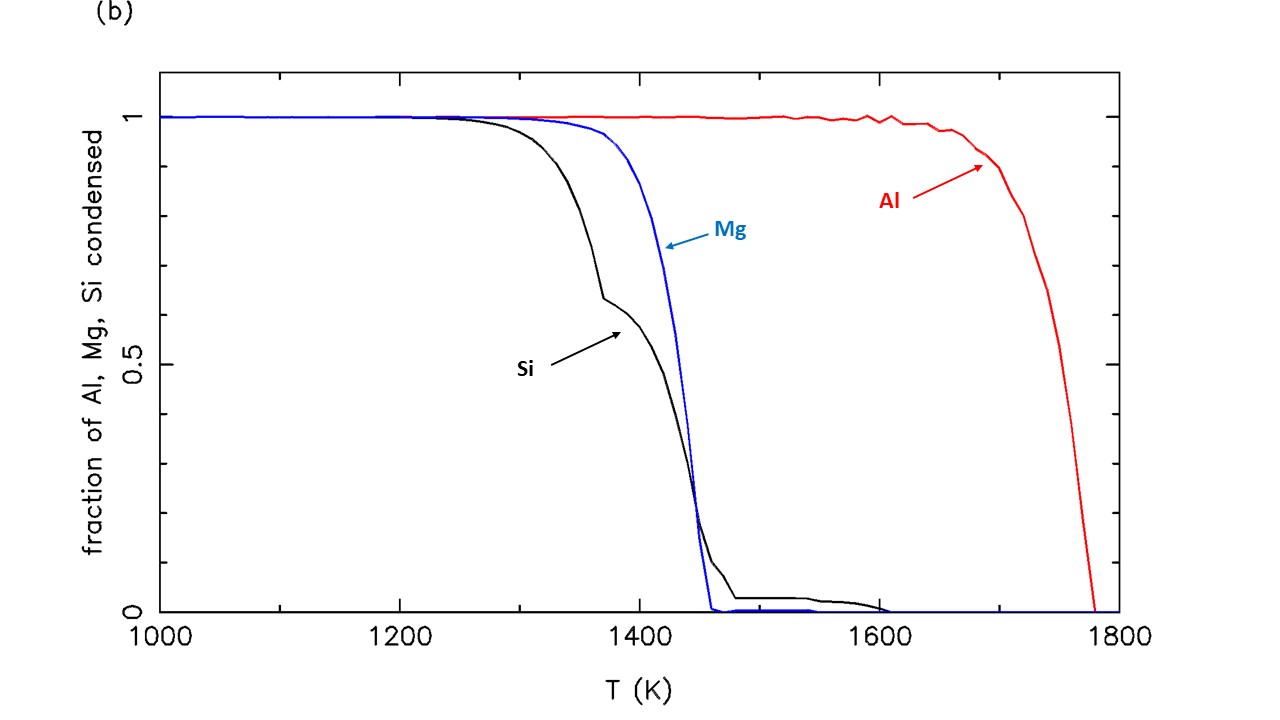}}
\caption{\linespread{0.5}\footnotesize a) Fraction of the total condensable mass from a cooling vapor of solar composition (i.e., CI composition, including volatiles) at $P_{tot} = 10^{-3}$ bar. Arrows indicate the onset of pure solid phase condensation together with their temperature. Abbreviations, cor : corundum (Al$_2$O$_3$) ; hib : hibonite (CaAl$_{12}$O$_{19}$) ; per : perovskite (CaTiO$_3$); grs : grossite (CaAl$_4$O$_7$) ; mel : mellitite (Ca$_2$Al$_2$SiO$_7$) ; fas : Aluminous-rich clinopyroxene (CaAl$_2$SiO$_6$ -- mimicking fassaite-like Fe poor, Ca, Al, Ti pyroxene); sp : spinel (MgAl$_2$O$_4$) ; met : Fe-Ni metal ; fo : forsterite (Mg$_2$SiO$_4$) ; an : anorthite (CaAl$_2$Si$_2$O$_8$) ; en : enstatite (MgSiO$_3$) ; alb : albite (NaAlSi$_3$O$_8$). b) Fractions of available Al, Mg and Si condensed as a function of temperature. Note that Al/Si and Mg/Si ratios of condensates become solar at a temperature of $\sim 1,250$K. Note also that the Al/Mg ratio becomes solar at larger temperature than the Al/Si ratio.}
\label{NewFig}
\end{figure}

Figure~\ref{NewFig} shows the fractions of the condensable mass and of Al, Mg, and Si that have condensed as a function of temperature, respectively. The comparison with published results on the condensation of a gas of solar composition (Yoneda and Grossman, 1995; Ebel and Grossman, 2000; Ebel 2006) shows a good agreement in terms of both the computed condensation temperatures and the sequence of appearance of solid phases upon cooling (Fig.~\ref{NewFig}a), despite the use of only pure solid phases. 

As is well known, corundum starts to condense first at a temperature of $\sim 1,775$~K at the pressure considered here ($10^{-3}$ bar). It is followed by calcium aluminates (hibonite, perovskite, grossite and melilite), which set the Al contents of the first condensates to be very high. Magnesium and Si start to condense later, almost simultaneously at $T \sim 1,600$~K with the onset of the condensation of melilite. Both Mg and Si contents  increase as cooling proceeds in response to condensation of fassaite and spinel at lower temperature. Initially, Si exceeds Mg in terms of fraction condensed but then, at 1,250~K$< T <$1,450~K, Mg exceeds Si because of the condensation of forsterite (Mg$_2$SiO$_4$), the first phase to remove major fractions of Si and Mg from the gas. Further cooling of the solar gas promotes the condensation of enstatite ($\sim 1,370$~K) and leads to the subsequent increase in Si of the solid. By the time the temperature has decreased to $\sim 1,250$~K, all available Al, Mg and Si atoms have condensed, which implies that the solar composition in terms of Al/Si and Mg/Si is achieved at this temperature.

\begin{figure}[t!]
  \centerline{\includegraphics[height=6.cm]{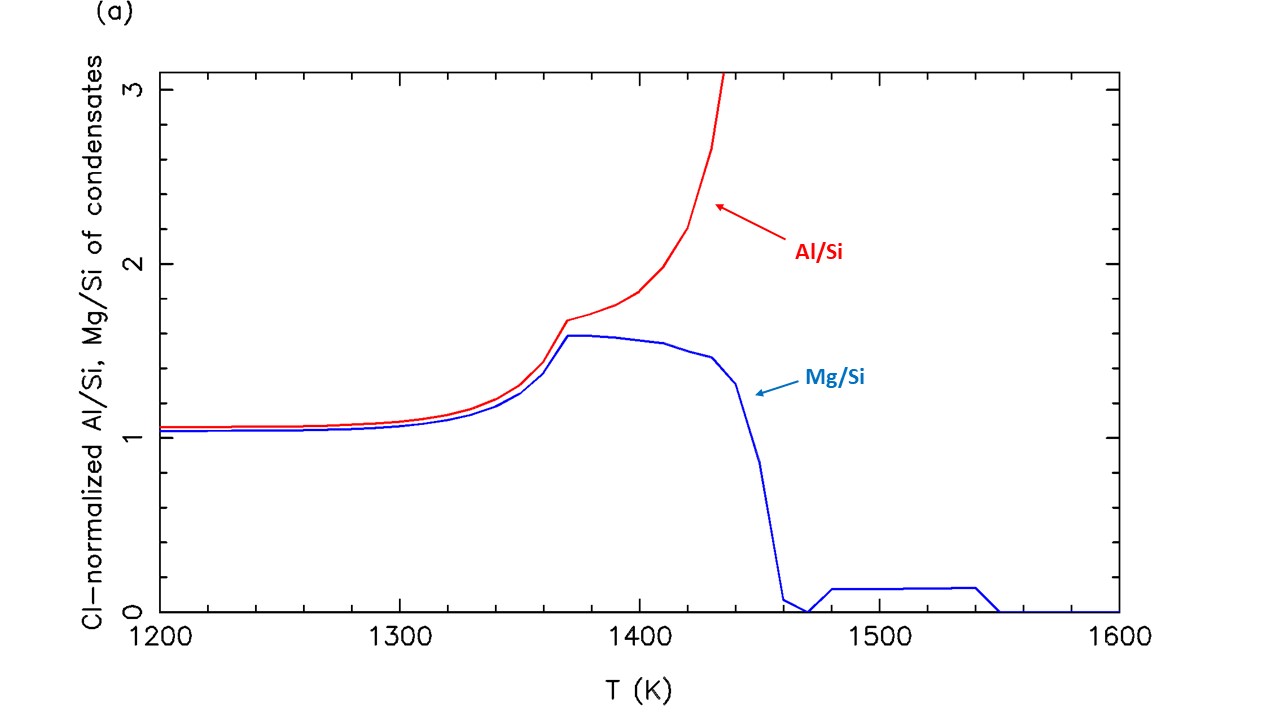}}
  \centerline{\includegraphics[height=6.cm]{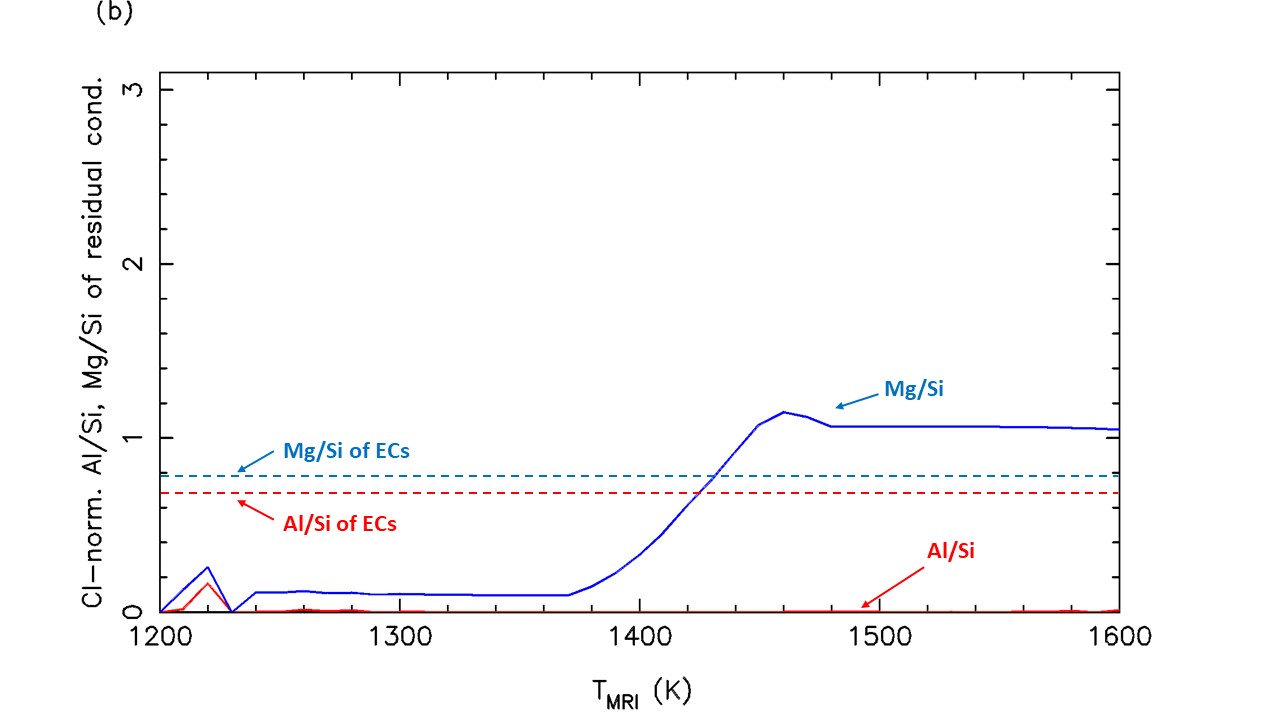}}
\caption{\linespread{0.5}\footnotesize Top panel: the solid curves show the Al/Si (red) and Mg/Si (blue) ratios (normalized to CI values) of the condensed material as a function of gas temperature. The arrow indicates that the Al/Si ratio increases above the upper limit of the plot for $T>1,430$K. 
Bottom panel: the same but for the residual condensates, assuming that all the early condensed material has been sequestered in planetesimals at the temperature $T_{MRI}$. The horizontal lines show the mean Al/Si and Mg/Si ratios for ECs.}
\label{ratios}
\end{figure}

The top panel of Fig.~\ref{ratios} shows the resulting Al/Si (red) and Mg/Si ratios (blue) as a function of temperature, normalized to the respective ratios in CI meteorites (i.e., the solar composition ratios).  The Al/Si ratio drops monotonically with decreasing temperature and reaches the solar ratio (i.e., one when normalized) at $T\sim 1,250$~K. The Mg/Si ratio instead starts sub-solar but then becomes supra-solar for 1,250~K$<T<1,450$~K, with the bump due to the condensation of forsterite. 

If condensates are isolated from the gas at a temperature $T_{MRI}$, they maintain the composition of the solids condensed at that temperature; thus, their Al/Si and Mg/Si ratios are those shown in Fig.~\ref{ratios}a at $T=T_{MRI}$. Strictly speaking, isolation occurs when condensates grow big enough to prevent further gas-solid exchanges. For example, given a diffusion coefficient of about $10^{-16}$m$^{2}$/s at 1,400~K for Fe-Mg interdiffusion in olivine (Holzapfel et al., 2003), a mm-size grain would equilibrate in approximately 300~y and a cm-size grain in 30,000~y. However, according to aggregation experiments it is likely that most grains are sub-mm in size (G\"uttler et al., 2010). Thus the most effective way to isolate solid material from exchange reactions with the gas is to incorporate it into macroscopic planetesimals, as explained in the previous section. 

Consider now the possibility that at some temperature $T_{MRI}$ the condensed solids are sequestered in planetesimals. Then, with further decreasing temperature, new solids will condense from the residual gas. The bottom panel of Fig.~\ref{ratios} shows the Al/Si and Mg/Si ratios of these residual condensates as a function of $T_{MRI}$. They are completely different from those of the first condensates. The Al/Si ratio of the residual condensates is strongly sub-solar (essentially zero below 1600 K, when all Al is condensed) and the Mg/Si ratio is also sub-solar for $T_{MRI}<1,450$~K. Figure~\ref{ratios}b shows, for comparison, the Al/Si and Mg/Si ratios of ECs. Note that at no value of $T_{MRI}$ do the residual condensates have simultaneously both the Al/Si and Mg/Si ratios of these chondrites.

\subsection{Making non-carbonaceous chondrites} \label{results}

It is dynamically unlikely and chemically impossible that the NCCs formed solely from residual condensates. {A fraction of the first condensed grains may have been small enough to avoid piling up at $T_{MRI}$, thus remaining at equilibrium with the cooling gas. These grains would have eventually acquired a solar composition in terms of concentrations of Al, Mg and Si and -possibly also- moderately volatile elements. In addition, other grains with solar {elemental} composition may have arrived from farther out in the disk. We stress that these grains did not necessarily have the CI content of water and of other highly volatile components, condensible only at low temperature\footnote{As we will see below, 60--90\% of solar-composition material is required in the NCCs and the Earth. If the solar-composition material had been real CI material, both these chondrites and the Earth would be rich in water and other volatile elements of comparable volatility, which is not the case.}. {Likewise, the isotopic properties of this material are not necessarily the same as for CI meteorites.}  Thus, hereafter when using the term "solar-composition" we {simply refer to material characterized by solar abundances of refractory and moderately volatile elements}. We postulate that the  NCCs formed as a second generation of planetesimals from a mixture of these solar-composition grains and residual condensates, so that they appear as if a refractory-rich component had been partially subtracted, as the data suggest.}

\begin{figure}[t!]
\centerline{\includegraphics[height=7.cm]{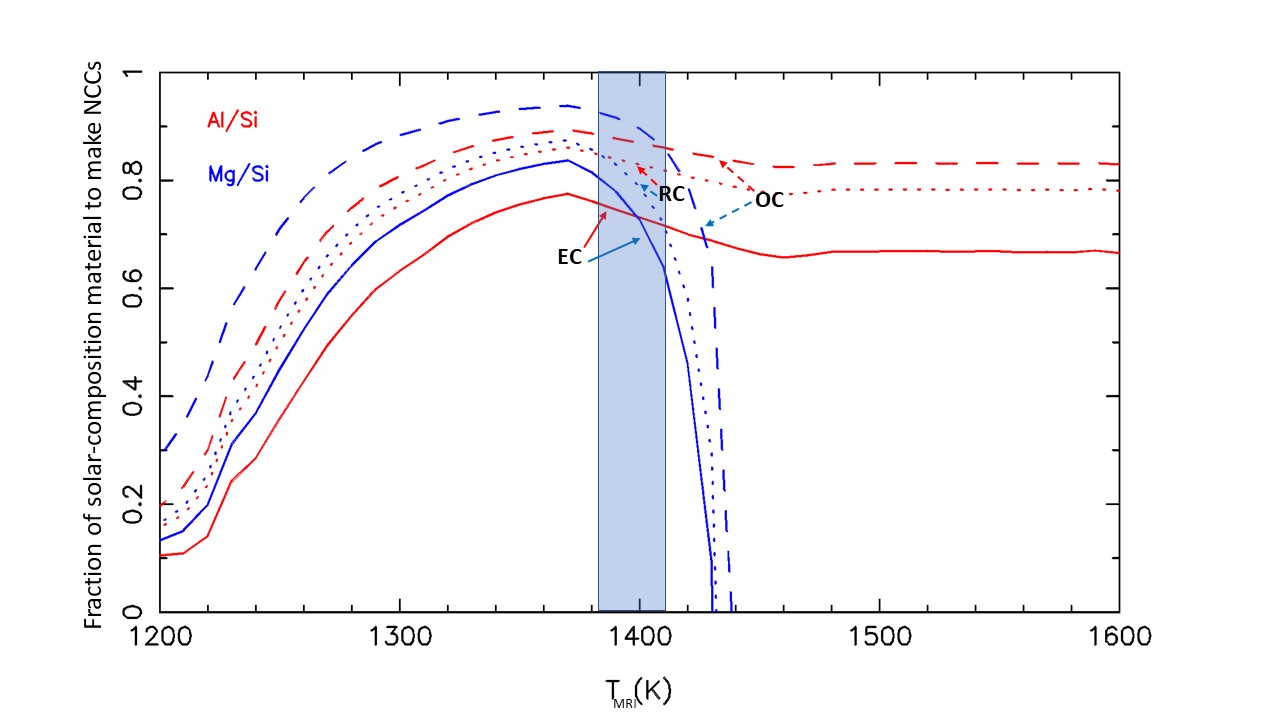}}
\caption{\linespread{0.5}\footnotesize The fraction of solar-composition material that needs to be mixed with residual condensates to reproduce the Al/Si (red) and Mg/Si (blue) ratios of ECs (solid), OCs (dashed) {and RCs (dotted)}, as a function of the temperature of formation of the first planetesimals, $T_{MRI}$. For ECs and OCs {we use the mean of the ratios of the various sub-classes}: Al/Si=0.67 and Mg/Si=0.78 and Al/Si=0.83 and Mg/Si=0.91 respectively, relative to CI ratios. The vertical band shows the range of values of $T_{MRI}$ that satisfy the considered elemental ratios of all the NCCs types: 1,385--1,410 K.}
\label{ECs}
\end{figure}

{Thus, for each temperature $T_{MRI}$ we compute the relative amount of solar-composition material $M_{Sol}$ that has to be mixed with the residual condensates to reproduce the Al/Si and/or the Mg/Si ratio of the NCCs (Fig.~\ref{ECs}).} $M_{Sol}$ is computed from the two equations:
\begin{eqnarray}
  R^{NCC}&=&{{f_{Al,Mg}^{res}M_{res}+\tilde f_{Al,Mg}^{CI} M_{Sol}} \over{f_{Si}^{res}M_{res}+\tilde f_{Si}^{CI} M_{Sol}}}\cr
  1&=&M_{res}+M_{Sol}
\label{eq.MCI}
\end{eqnarray}
where  $R^{NCC}$ is the Al/Si or the Mg/Si weight ratio in {the considered NCCs (ECs, RCs or OCs)}; $f_{Al,Mg}^{res}$ and $f_{Si}^{res}$ are the concentrations by mass of Al, Mg and Si in residual condensates (all these quantities depend on $T_{MRI}$ and are shown in Fig. S1), whose total mass is $M_{res}$; $\tilde f_{Al,Mg}^{CI}$  and $\tilde f_{Si}^{CI}$ are 1.66 times the concentrations of Al, Mg and Si in CI meteorites (see table~S1 for element fraction in meteorites). The factor 1.66 takes into account the lack of volatiles (e.g., water) in the solar-composition material that we consider compared to CI meteorites, which reduces their total mass to 60\%.

Obviously, for each meteorite class the concordia situation occurs at the intersection between the two solutions, when both the Al/Si and Mg/Si ratios are reproduced for the same combination of $T_{MRI}$ and $M_{Sol}$. {Note that the solutions for $T_{MRI}$ are {very similar (1,385--1,405~K) for all the NCC classes}. This means that the residual condensate material (or, equivalently, the subtracted refractory component) is the same for these classes of meteorites and that the different bulk chemical compositions of the ECs and OCs are mostly due to different mixing proportions with the solar-composition material.} {Notice that the relative mass of solar-composition material exceeds in all cases the fraction of matrix in these meteorites. This means that the formation of chondrules and matrix occurred after mixing solar-composition material and residual condensates. Hence, it does not contradict our model that the RCs are intermediate between ECs and OCs in terms of proportion of solar-composition material even if they have the largest proportion of matrix.}

{The elemental and the mineralogical compositions of the refractory material sequestered at 1,400~K are shown in the top panels of Fig.~\ref{camembert}}. {We have repeated the calculation for a pressure of $10^{-4}$ bar. Conceptually the results are the same, but all condensation temperatures shift to lower values; $T_{MRI}$ is found to be in the range 1,310--1,330~K.}

\section{Other elements}
\label{others}

{Now that $T_{MRI}$ is determined, we can check the consistency of the model with other {elements with different condensation temperatures.}

{We start our analysis with a moderate refractory element like Fe. For $P=10^{-3}$~bar, the residual condensates would have a sub-solar Fe/Si ratio (Fig. S1). However, for $P=10^{-4}$ bar  Fe is less refractory and the Fe/Si ratio at $T_{MRI}$ is basically unfractionated.}  The amount of Fe varies widely within the ECs and OCs, from EH to EL and H to LL. These changes cannot be explained by our simple model here. They may require different oxidation conditions, in the disk, during chondrule formation (Larimer, 1979; Alexander, 2019). We notice that because Fe tends to be present in meteorites as distinct metal blebs, Fe enrichment and depletion may also be due to size-sorting effects during the streaming instability.} 

\begin{figure}[t!]
\centerline{\includegraphics[height=7.cm]{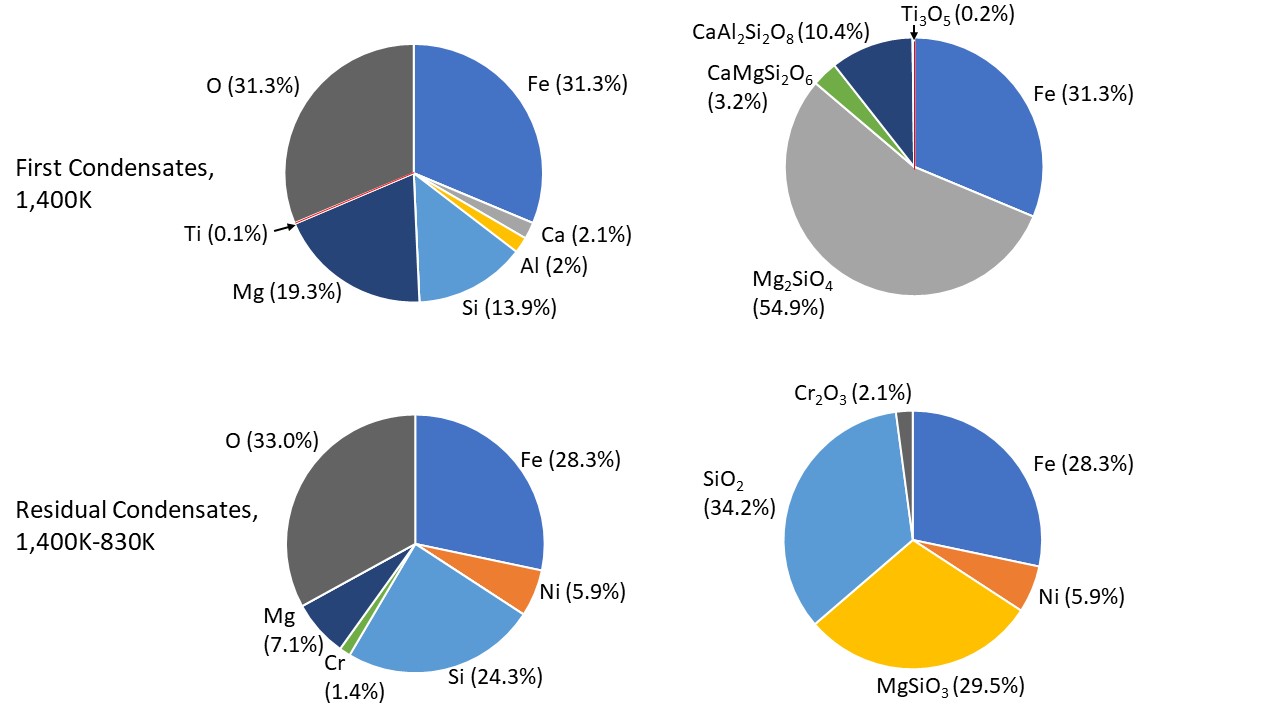}}
\caption{\linespread{0.5}\footnotesize The elemental (left pie-charts) and mineralogical (right pie-charts) weight fractions for the first condensates at 1,400K (top) and residual condensates down to 830K (bottom).}
\label{camembert}
\end{figure}

{Notice that in the top panels of Fig.~\ref{camembert} Ni does not appear. This is an artifact of our calculation scheme since only pure Fe and pure Ni metal phases (and not the single FeNi solid solution) have been taken into account. Thus, the Ni/Fe ratios in both the first and the residual condensates are not realistic.}  

{Sodium is a more interesting case, because in the NCCs, with the exception of the EH meteorites, it shows a striking correlation with Al (see Fig.~\ref{Palme}).} The same is true for the other alkali elements (Alexander, 2019). According to our model, when the refractory material is sequestered at 1,400~K, Na has not yet condensed, so the Na-Al correlation is not expected. 

{However, as pointed out by Barshay and Lewis (1976), the sequestration of refractory elements prevents their further reaction with the gas at lower temperatures, making Na less able to condense, i.e., more volatile. For instance, in an equilibrium condensation sequence, {at $P=10^{-3}$~bar,} the condensation of Na would start by forming NaAlSi$_3$O$_8$ (albite) at 1,150~K. But as Al has been removed from the gas, this mineral cannot form. The first refractory-element-free condensate that Na can form is Na$_2$S (Fegley and Lewis, 1980), which starts to condense below 830~K at $P=10^{-3}$~bar. Thus, if gas condensation ``ends'' before this temperature is reached the residual condensates do not contain Na and the final depletion in Na in the ECs and OCs correlates perfectly with the depletion in Al, because both Na and Al are acquired only from the solar-composition component of these meteorites.  

  We stress that {it is natural that} the condensation sequence ends {when there is still gas left in the system. In fact, condensation requires cooling the gas but, although the disk cools over time at each location, parcels of gas are also radially transported towards the star. Thus, there is a competition between heating of gas due to its inward radial drift and secular cooling of the disk. Morbidelli et al. (2016) showed that a viscously evolving disk transitions from condensing (cooling) gas parcels to inwardly drifting (heating) parcels after half a viscous timescale} ($r^2/\nu$), which is a small fraction of the disk's lifetime. {In the disk model of Bitsch et al. (2015) this happens at 1~AU at $t\sim 10^4$~y, when the stellar accretion rate is $10^{-6}M_\oplus$/y, before the temperature has reached $\sim 800$~K.} Assuming this is the case, we show in the bottom row of Fig.~\ref{camembert} the elemental and the mineralogical compositions of the residual condensates.}

{The case of Cr is more difficult. Still for our nominal case of $P=10^{-3}$~bar 50\% of Cr condenses at 1,300K, too close to $T_{MRI}$ to invoke the premature end of the condensation sequence, but not high enough to remove most of the Cr with the first condensates. Thus, we should expect that the residual condensates were enriched in Cr and that the NCCs have Cr/Si higher than the CI value. Instead, the depletion of Cr correlates with that of Al (Alexander, 2019).  Accounting for the solubility of Cr in Fe-metal and olivine allows for the removal of $\sim$50\% of the Cr with the first condensates, but $\sim$50\% of Si is also removed, so that the residual condensates should be at best un-fractionated in Cr/Si, which is insufficient to explain the observed Al-Cr correlation. However, we find that if the C/O ratio of the gas is increased to 0.9, $T_{MRI}$ decreases to 1,280K (because the condensation temperatures of Al, Mg and Si decrease), so that nearly 100\% of Cr is bound in the first condensates if one accounts for its solubility in iron and olivine. Thus a C/O$\ge 0.9$ enables the Cr-Al correlation in the NCCs to be explained.  The same is true at lower pressure. A similar C/O ratio has been invoked by several authors (e.g., Lodders and Fegley, 1993) as the sole way to explain various properties of enstatite chondrites.  It is interesting that from a completely different approach we also reach this conclusion, with the difference that in our model the high C/O ratio affected the formation of residual condensates that are a component of all the NCCs, not just of the ECs (although more prominent in ECs). The equivalent of Fig.~\ref{camembert}, except for C/O=0.9, is given in Fig. S4.}

{Sulfur is more volatile than Na in a gas with solar C/O. But it starts to behave as a refractory element for C/O$> 0.9$, when oldhmite (CaS) begins to condense. For C/O$=1.2$, 40\% of S is condensed at $T_{MRI}$ ($=$1,130~K for $P=10^{-3}$~bar). No S condenses from the residual gas before Na. Thus, S and Na depletions should correlate in the NCCs, but such a correlation is not observed. Sulfur is mostly found in the matrix and high concentrations of S are only observed in EC chondrules. Remembering that the mixing between residual condensates and solar-composition material should be done at the level of chondrule precursors this observation suggests that, for some reason, S was lost in the formation of OCs and RCs chondrules, but not Na (Alexander et al., 2008). } 
 
\section{The Earth}
\label{Earth}

{If the first-formed planetesimals had accreted with each other and possibly with additional solar-composition material, they would have formed a planet enriched in refractory elements. It is therefore important to check whether this scenario could explain the supra-solar Al/Si and Mg/Si ratios of the Earth. 

{For simplicity we consider our nominal case of $P=10^{-3}$~bar and a solar C/O ratio.} We assume that the composition of the first-formed planetesimals is that obtained from the equilibrium condensation at $T_{MRI}=1,400$K (Fig.~\ref{camembert}, top - see also Fig. S2) and use eq. (1) again. The  Al/Si and Mg/Si ratios of the Earth increase as the relative mass fraction of these planetesimals incorporated in the planet increases (the rest of Earth's mass having a solar composition) and we find that ratios consistent with the observed values within their broad uncertainties can be obtained for a mass fraction of about 35-45\% (Fig.~\ref{terrestrial}).\footnote{We note that combining first planetesimals with ECs would lead to equivalent results. In fact, the combination of first-condensed material with its complement of residual condensates gives by definition material with a solar bulk composition. Thus, the $x$-axis of Fig.~\ref{terrestrial} can be simply interpreted as the excess of first-condensed material in the Earth, whatever the combination of material that is envisioned.}

\begin{figure}[t!]
\centerline{\includegraphics[height=7.cm]{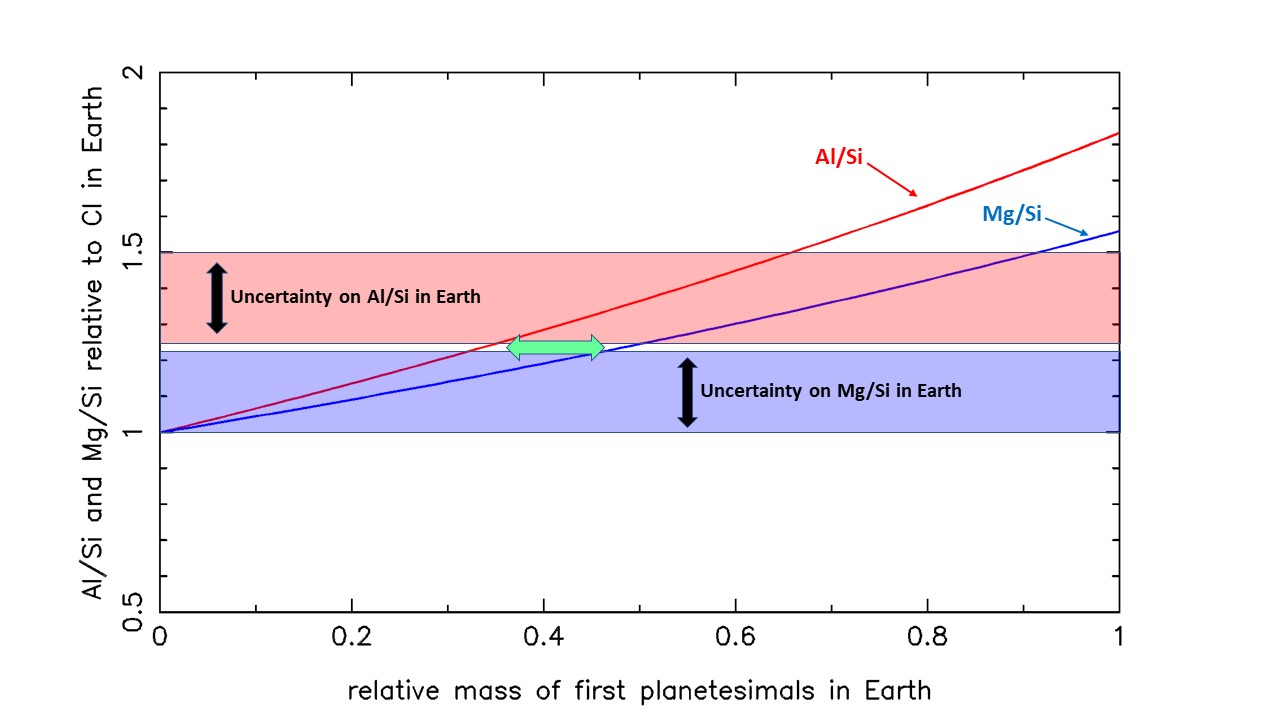}}
\caption{\linespread{0.5}\footnotesize The Al/Si (red) and Mg/Si (blue) ratios as a function of the mass fraction of first, refractory rich planetesimals contributing to the Earth, the rest being accreted from solar-composition material. The horizontal colored bands show the estimated bulk terrestrial Mg/Si and Al/Si ratios and their uncertainties. These are reproduced if the Earth comprises 35-45\% by mass of first planetesimals. }
\label{terrestrial}
\end{figure}

Alexander (2019) explicitly excluded the possibility that the refractory material missing from the NCCs (i.e., our first-formed planetesimals) has been accreted by the Earth. His conclusion is motivated by three arguments, which we review and discuss here. }

{First, he pointed out that the addition to solar-composition material of the refractory material missing from the NCCs does not fit well the Earth's composition. This is visible also in Fig.~\ref{terrestrial} because the Al/Si and Mg/Si ratios can be  simultaneously reproduced only at the opposite extremes of their error bars. If the uncertainties on these ratios are entirely due to the amount of Si incorporated in the terrestrial core, they are not independent. So, in this case one cannot claim success if one ratio is reproduced at the high-end of its uncertainty limit and the other at its low-end. However,  {Young et al. (2019) argued that the Earth or its precursors lost by evaporation about 12\% of Mg and 15\% of Si. This would have left the Mg/Si ratio of the Earth basically unchanged, but would have increased the Al/Si ratio by 17\%. For instance, in Fig.~\ref{terrestrial} at a value of 0.3 on the x-axis, the Mg/Si ratio predicted by our model falls in the middle of the observed uncertainty. The Al/Si ratio predicted by the model is too low (1.2), but with 15\% Si evaporation it would have increased to 1.38, i.e. again in the middle of observed uncertainty. Thus, our model is consistent with the Al/Si and Mg/Si ratios of the bulk Earth, even if their uncertainties are correlated, provided that the evaporation of Si and Mg invoked to explain their isotopic fractionation is taken into account.}   

Second, Alexander (2019) noted a correlation between the depletion of Al and the depletion of alkali elements in the NCCs. He concluded that alkalis were sequestered in the refractory material subtracted from these meteorites. Thus, if this material had been added to the Earth, our planet would be alkali-rich instead of alkali-poor. Although we have based our argument only on Na, in Sect.~\ref{others} we showed that the removal of a refractory component makes alkalis much more volatile than expected in an equilibrium condensation. Thus, the correlation between the depletions of alkalis and Al can be explained without the sequestration of the alkalis in the refractory material. Consequently, the addition of the refractory material to the Earth would make our planet alkali-poor, not alkali-rich. 

Finally, Alexander (2019) noted a correlation between the depletion of Al in the NCCs and the isotopic ratios of some elements, such as Ti and Cr. He concluded that the removed refractory material carried strong isotopic anomalies for these elements. However, these anomalies are not observed in the Earth, implying that our planet did not accrete the refractory material missing from the NCCs. We find (Fig.~\ref{camembert}a) that the refractory material at 1,400~K would have sequestered virtually all of the Ti and, {for a sufficiently high C/O ratio, all the Cr as well}. Thus, the observed differences in Ti and Cr isotopic compositions among the NCCs cannot be related to the condensation and removal of this refractory material and instead have to be due to the isotopic heterogeneity of the solar-composition material in the disk, although the origin of this heterogeneity is not well understood.

{ We remark that in the case of C/O$\ge 0.9$, the bulk Earth should have a Cr/Si ratio larger than the CI value (as Cr is captured in the first-formed planetesimals). But Cr is a moderate siderophile element, so that most of it could be in the Earth's core (e.g., McDonough and Sun, 1995), explaining why the bulk {\it silicate} Earth is depleted in Cr.} 

We end this discussion by commenting that it is certainly possible to envision that the refractory material sequestered from the NCCs disappeared without contaminating the Earth. For instance, refractory elements could be locked-up in grains large enough to be out of equilibrium with the gas, but still small enough to migrate by gas-drag into the Sun. But then, one has to invoke that the Earth accreted some other refractory-rich material, that did not contaminate the NCCs although being isotopically identical to the ECs. Possibly by lack of imagination, we cannot envision a reasonable scenario for this to have happened. Thus, we conclude that the addition to the Earth of the refractory material sequestered from the NCCs remains the best option.}

\section{Conclusions and discussions}
\label{Conclusions}

\subsection{Summary of results}

In this work, we have {proposed an astrophysical scenario, consistent with current models of disk evolution and planetesimal formation, that can explain the sequestration of refractory elements from the source region of the non-carbonaceous (NCC) chondrites and, therefore, their sub-solar Al/Si and Mg/Si ratios. 

In brief,} while the disk was cooling, the massive condensation of olivine changed the disk from a low-dust to a high-dust environment, {quenched the magneto-rotational instability of the disk, triggering} the sudden formation of a first generation of planetesimals and  the sequestration of the already condensed solid material. That material was, therefore, no longer available for the evolution of the gas-solid equilibrium chemistry. With a further decrease in temperature, new solids could form from residual gas. We called these ``residual condensates'' and the region where these events happened the ``residual condensate region'' (RCR). For genetic reasons, the first planetesimals and the residual condensates are complementary relative to the solar (CI) values in terms of Al/Si and Mg/Si ratios. 

{We envisioned that the NCCs formed from a combination of these residual condensates and material that had solar composition in terms of refractory and moderately volatile elements. This material consisted of local grains that were not incorporated in the first planetesimals and remained at equilibrium with the gas and -possibly- also grains that} formed farther out in the disk, and migrated into the RCR without suffering element fractionation. 

We have shown that, {if the first planetesimals formed at a temperature of $\sim 1,400$~K,} {(for a pressure of $10^{-3}$~bar and solar C/O, and a lower temperature, nevertheless exceeding 1,000~K, for lower pressure and/or increased C/O)} the mixture of appropriate proportions of solar-composition grains and residual condensates can reproduce simultaneously the Al/Si and Mg/Si ratios of all the NCCs. Obviously, the amount of solar-composition material has to increase from the ECs to OCs, as the latter have compositions that are closer to the CI ratios.  {We also explained the correlation between the deficits of Na and Al in the NCCs, as a result of the fact that the sequestration of Al makes Na much more volatile than in an equilibrium condensation sequence (see Sect.~\ref{others}).} {To explain the correlation between the deficits of Cr and Al, we had to invoke that the first condensates formed in a gas with C/O$\ge 0.9$, in agreement with previous work on the ECs (e.g., Larimer, 1975; Lodders and Fegley, 1993).}

We showed that the Al/Si and Mg/Si ratios of the Earth can also be understood if our planet formed from a mixture of first planetesimals and solar-composition material. {Alexander (2019) explicitly excluded this possibility, but we provided in Sect.~\ref{Earth} some possible solutions to the obstacles that he discussed for this scenario.}
Our full scenario is sketched in Fig.~S3, which presents a global view of the composition of the disk as time evolves.

Our model is appealing in that it solves two long-standing problems. First, it explains the sub-solar Al/Si and Mg/Si ratios of the NCCs within a fractionated condensation sequence. Second, it clarifies the genetic relationship between the Earth and the ECs, as discussed below. Nevertheless, our model is simple and should be regarded mostly as a proof of concept. For instance, we assumed that all first planetesimals formed at the same temperature, {so that all the objects we considered are composed of only two components: residual condensates and solar-composition material for the NCCs or refractory-rich planetesimals and solar-composition material for the Earth.} 

\subsection{Earth -- enstatite chondrites relationship}

As discussed in the introduction, the isotopic compositions of the Earth and of the ECs in terms of non-mass dependent isotopic variations are extremely similar. No other known meteorite class approximates the Earth better than the ECs from the isotopic point of view. But element ratios are so different that it is {difficult to envision} making the Earth out of the ECs. Here we find that an important component of the Earth is represented by the first planetesimals, while an important component of the ECs is represented by residual condensates. First planetesimals and residual condensates are complementary in terms of element ratios, but they both form out of the same gas in the RCR, so it is not surprising that they are isotopically very similar. In some sense, with the first refractory-rich planetesimals we have identified the hidden reservoir of objects with the same isotope composition as the ECs but different chemical composition, that was postulated by Dauphas (2017) to explain the Earth. {The opposite mass-dependent isotopic fractionation of Si in the Earth and ECs may be a natural consequence of the formation of the first condensates, as discussed by Kadlag et al. (2019).}

Pushed further, the genetic relationship between the Earth and the ECs illustrated in this paper can also help to understand the small differences in their respective Nd, Mo and Ru isotopic compositions. The Earth has an endmember isotopic composition relative to known meteorites, slightly enriched in isotopes produced by the s-process with respect to the ECs (see Burkhardt et al. 2016 and Bouvier and Boyet, 2016 for Nd; Burkhardt et al., 2014 for Mo and Fischer-G{\"o}dde and Kleine, 2017 for Ru). If, for some reason, the material that condensed directly in the RCR is enriched in  s-process isotopes relative to the solar-composition grains coming into the RCR once the temperature has decreased, we can understand this difference. In fact, in this case the first planetesimals and the residual condensates would have the same proportion of  s-process isotopes but, as shown in Sect.~\ref{results}, the Earth, EC and OC meteorites incorporated increasing fractions of solar-composition material, thus explaining -at least qualitatively- the relative ranking of  s-process enrichment/deficiency observed in these objects.

\subsection{Chondrules and CAIs}

Although we invoke here the mixing of residual condensates and solar-composition grains to make the ECs and OCs, these meteorites are made of chondrules and not grains. Chondrule formation is still not fully understood (Krot et al., 2018) and presumably the mixing we invoke occurred at the level of chondrule precursors and were transformed in the high-temperature events that led to chondrule formation. Our elemental and isotopic considerations should nevertheless hold even in this more complex scenario. 

The relationship between the first condensed material discussed in this paper and the refractory materials (CAIs, AOAs) found in mostly in the CCs is more elusive. It would be tempting to identify CAIs with a fraction of the minerals condensed at $T\gtrsim 1,500$~K that escaped incorporation in the first generation of planetesimals and somehow reached the outer disk. But CAIs also carry isotopic anomalies (for instance in Nd or Mo) related to the p- and  r-processes, which are not observed in the Earth, ECs and OCs. {However, Ebert et al. (2018) claim to have identified a refractory component isotopically distinct from CAIs in OCs.  The first condensed material discussed in this paper would rather correspond to this refractory component rather than CAIs.}

\subsection{Protoplanetary disk properties and grain dynamics}

Although our work does not specify where the RCR was located, the fact that the first planetesimals have to be a primary component of the Earth suggests that the sequence of events described in this paper happened at $\sim 1$~AU. This implies that the disk at  $\sim 1$~AU was originally hotter than 1,400K. Viscous-disk models (Bitsch et al., 2015) suggest that the temperature at 1~AU was initially lower, but these models neglect the heat released by the accretion of gas from the interstellar medium onto the disk, which was presumably vigorous at early times (P. Hannebelle, private communication). So, our model does not seem unreasonable, but it points to a disk much more complex than those usually envisioned.

{As long as the infall of gas on the inner disk is vigorous, the temperature remains very high. For instance, Bailli\'e et al. (2018) find a temperature larger than 1,500~K up to 2~AU from the Sun for the first $10^5$~y. Thus, in this timeframe, only the gas viscously spreading beyond 2 AU would cool and condensate. Then, when the infall wanes, the temperature in the inner disk starts to decrease and condensation can occur {\it locally}, producing a second condensation front moving towards the star. The fractional condensation discussed in this paper would correspond to this second condensation front.  This view of two condensation fronts could explain the difference between CAIs and the non-CAI refractory component of Ebert et al. (2018), if the isotopic properties of the infalling gas changed over time, as invoked in Nanne et al. (2019) to explain the isotopic radial heterogeneity of the protosolar disk\footnote{see Jacquet, 2019, for a discussion of the possibility that CAIs formed relatively far from the Sun from the point of view of $^{10}$Be production}. Similarly, if the gas changed of chemical composition, the high local C/O ratio required to explain Cr-abundances can be explained as well. Indeed, filaments of gas are observed to fall onto protoplanetary disks (Alves et al., 2019), carrying a different C/O composition (Segura-Cox, private communication).  Moreover,} the rapid drop of temperature to $\sim 850$~K followed by slow cooling, that our model requires to end the residual condensation before the condensation of Na, could be due to the transition from a disk dominated by a rapidly waning infall of material from the ISM to a disk dominated by its own viscous evolution. {Admittedly, this scenario is speculative, but it is not implausible}.

{Our model also implies that the transition from the MRI-active to the MRI-dead zones of the disk happened at a temperature between 1,060 and 1,400~K depending on pressure and C/O ratio, but nevertheless larger than the temperature of condensation of K, whereas Desch and Turner (2015) argued that the condensation of K is the key to reducing the ionization of the gas and to quenching the MRI. The temperature $T_{MRI}$ we obtained was constrained by the elemental ratios in the NCCs. If the first planetesimals could form only in a MRI-inactive disk and this requires the condensation of K, we expect that K and refractory elements should always be correlated, which is not the case for the Earth or the CCs. A possible solution of this conundrum is that the first planetesimals did not form at the MRI-active/inactive transition, but in the turbulent disk, by concentrating dust into vortices.}

In our model, the residual condensates survive in the disk and mix with solar-composition grains to form eventually the NCCs. However, the parent bodies of these meteorites formed relatively late, after about 2-3 My (Sugiura and Fujiya, 2014). How grains could survive in the disk for so long despite their tendency to migrate towards the Sun is an open problem that this work does not help to solve. Possible explanations involve turbulent diffusion counteracting radial drift near the midplane of the disk (Ciesla, 2007), reduced or reversed radial drift due to the partial depletion of gas in the inner disk (Ogihara et al., 2018), recycling of material in disk winds or jets (Ciesla, 2009), {lock-up in planetesimals and later release as collisional droplets (Johnson et al., 2015).} We stress that in our model we do not need that all residual condensates survive. {If the total mass of planetesimals with NCC-like compositions was small, only a tiny fraction of residual condensates may have survived.}

\subsection{Closing remarks}

In summary, up to now the greatly different ratios among major refractory elements in the Earth and in non-carbonaceous chondrites have challenged our understanding of the formation of inner Solar System bodies. Although several unknowns remain, we have shown that the accretion of a first generation of planetesimals during the condensation sequence of refractory elements and the consequent formation of residual condensates are key processes that contribute to the solution of this problem.

\section{Acknowledgments}

We thank K. Burkhardt, T. Kleine, E. Young and N. Dauphas for valuable suggestions on an earlier version of this manuscript. {We thank in particular C. Alexander for an open discussion on our respective models and B. Fegley and K. Lodders for an in depth critical assessment of our results during their visit to Nice and their historical perspective on this long-studied subject. The reviews from two anonymous referees have also helped to improve the manuscript and clarify the results.}

\section{References}

\begin{itemize}

\item[--] Alexander, C.~M.~O'D., Grossman, J.~N., Ebel, D.~S., Ciesla, F.~J.\ 2008.\ The Formation Conditions of Chondrules and Chondrites.\ Science 320, 1617.

\item[--] Alexander, C.~M.~O'D.\ 2019.\ Quantitative models for the elemental and isotopic fractionations in the chondrites: The non-carbonaceous chondrites.\ Geochimica et Cosmochimica Acta 254, 246.

\item[--] Alves, F.~O., and 6 colleagues 2019.\ Gas flow and accretion via spiral streamers and circumstellar disks in a young binary protostar.\ Science 366, 90.

\item[--] Bailli{\'e}, K., Marques, J., Piau, L.\ 2019.\ Building protoplanetary disks from the molecular cloud: redefining the disk timeline.\ Astronomy and Astrophysics 624, A93.
  
\item[--] Bale, C. W., et al., 2016. FactSage Thermochemical Software and Databases, 2010-2016, Calphad, vol. 54, pp 35-53, 2016

\item[--] Ballmer, M.~D., Houser, C., Hernlund, J.~W., Wentzcovitch, R.~M., Hirose, K.\ 2017.\ Persistence of strong silica-enriched domains in the Earth's lower mantle.\ Nature Geoscience 10, 236.

\item[--] Barshay, S.~S., Lewis, J.~S.\ 1976.\ Chemistry of primitive solar material..\ Annual Review of Astronomy and Astrophysics 14, 81.

\item[--] Bitsch, B., et al.\ 2015.\ The structure of protoplanetary discs around evolving young stars.\ Astron. Astrophys. 575, A28. 

\item[--] Bouvier, A., Boyet, M.,\ 2016.\ Primitive Solar System materials and Earth share a common initial $^{142}$Nd abundance.\ Nature 537, 399-402. 

%\item[--] Brennecka, G.~A., et al., \ 2018.\ Molybdenum Isotopic Evidence for a Distal Formation of Refractory Inclusions.\ LPSC  49, 2429. 

\item[--] Burkhardt, C., et al.,\ 2014.\ Evidence for Mo isotope fractionation in the solar nebula and during planetary differentiation.\ Earth Planet. Sci. Lett. 391, 201-211. 

\item[--] Burkhardt, C., et al.,\ 2016.\ A nucleosynthetic origin for the Earth's anomalous $^{142}$Nd composition.\ Nature 537, 394-398. 

%\item[--] Chatterjee, S., Tan, J.~C.,\ 2014.\ Inside-out Planet Formation.\ Astrophys. J. 780, 53. 

\item[--] Ciesla, F.~J.,\ 2007.\ Outward Transport of High-Temperature Materials Around the Midplane of the Solar Nebula.\ Science 318, 613. 

\item[--] Ciesla, F.~J.,\ 2009.\ Dynamics of high-temperature materials delivered by jets to the outer solar nebula.\ Met. Pl. Sci. 44, 1663-1673. 

\item[--] Dauphas, N., Poitrasson, F., Burkhardt, C., Kobayashi, H., Kurosawa, K.\ 2015.\ Planetary and meteoritic Mg/Si and ${{\delta}}^{30}$ Si variations inherited from solar nebula chemistry.\ Earth and Planetary Science Letters 427, 236.

\item[--] Dauphas, N.,\ 2017.\ The isotopic nature of the Earth's accreting material through time.\ Nature 541, 521-524.  

%\item[--] Davis, A.~M., Richter, F.~M.,\ 2014.\ Condensation and Evaporation of Solar System Materials.\ Met. Cosmochem. Process. 335-360. 

\item[--] Desch, S.~J., Turner, N.~J.\ 2015.\ High-temperature Ionization in Protoplanetary Disks.\ The Astrophysical Journal 811, 156.

\item[--] Ebel D. S, 2006. Condensation of Rocky Material in Astrophysical Environments. In: Meteorites and the Early Solar System II, D. S. Lauretta and H. Y. McSween Jr. (eds.), University of Arizona Press, Tucson, 943 pp., p.253-277.

\item[--] Ebel D. S and Grossman L., 2000. Condensation in dust-enriched systems. Geochim. Cosmochim. Acta, 64, 339-366.

\item[--] Ebert, S., et al.,\ 2018.\ Ti isotopic evidence for a non-CAI refractory component in the inner Solar System.\ Earth Planet. Sci. Lett.  498, 257-265. 

\item[--] Eriksson, G, Hack, K., 1990. ChemSage -  A computer program for the calculation of complex chemical equilibria, Metall. Trans. B, 21B, 1013 

\item[--] Fegley, B., Lewis, J.~S.\ 1980.\ Volatile element chemistry in the solar nebula: Na, K, F, Cl, Br, and P.\ Icarus 41, 439.

\item[--] Fischer-G{\"o}dde, M., Kleine, T.,\ 2017.\ Ruthenium isotopic evidence for an inner Solar System origin of the late veneer.\ Nature 541, 525-527.

\item[--] Flock, M., Fromang, S., Turner, N.~J., Benisty, M.\ 2017.\ 3D Radiation Nonideal Magnetohydrodynamical Simulations of the Inner Rim in Protoplanetary Disks.\ The Astrophysical Journal 835, 230.

\item[--] G{\"u}ttler, C., et al.,\ 2010.\ The outcome of protoplanetary dust growth: pebbles, boulders, or planetesimals?. I. Mapping the zoo of laboratory collision experiments.\ Astron. Astrophys. 513, A56. 

\item[--] Hezel, D.~C., et al.,\ 2008.\ Modal abundances of CAIs: Implications for bulk chondrite element abundances and fractionations.\ Met. Pl. Sci. 43, 1879-1894. 

%\item[--] Hin R. C., et al., 2017. Magnesium isotope evidence that accretional vapour loss shapes planetary compositions. Nature 549: 511-515

\item[--] Holzapfel, C., et al., 2003. Pressure effect on Fe-Mg interdiffusion in (FexMgx-1)O, ferropericlase. Phys. Earth Planet. Int. 139, 21-34.

\item[--] Hyung, E., Huang, S., Petaev, M.~I., Jacobsen, S.~B.\ 2016.\ Is the mantle chemically stratified? Insights from sound velocity modeling and isotope evolution of an early magma ocean.\ Earth and Planetary Science Letters 440, 158.

%\item[--] Ilgner, M., Nelson, R.~P.,\ 2006.\ On the ionisation fraction in protoplanetary disks. I. Comparing different reaction networks.\ Astron. Astrophys. 445, 205-222. 
 
%\item[--] Jacquet, E., Alard, O., Gounelle, M.\ 2015.\ The formation conditions of enstatite chondrites: Insights from trace element geochemistry of olivine-bearing chondrules in Sahara 97096 (EH3).\ Meteoritics and Planetary Science 50, 1624.
  
\item[--] Jacquet, E.\ 2019.\ Beryllium-10 production in gaseous protoplanetary disks and implications for the astrophysical setting of refractory inclusions.\ Astronomy and Astrophysics 624, A131.

\item[--] Javoy, M.,\ 1995.\ The integral enstatite chondrite model of the Earth.\ Geophys. Res. Lett. 22, 2219-2222.

%\item[--] Javoy, M., et al., 2010.\ The chemical composition of the Earth: Enstatite chondrite models.\ Earth Planet. Sci. Lett.  293, 259-268. 

\item[--] Johansen, A., Youdin, A.\ 2007.\ Protoplanetary Disk Turbulence Driven by the Streaming Instability: Nonlinear Saturation and Particle Concentration.\ The Astrophysical Journal 662, 627. 
 
\item[--] Johnson, B.~C., Minton, D.~A., Melosh, H.~J., Zuber, M.~T.\ 2015.\ Impact jetting as the origin of chondrules.\ Nature 517, 339.

\item[--] Kadlag, Y., Tatzel, M., Frick, D.~A., Becker, H.\ 2019.\ The origin of unequilibrated EH chondrites - Constraints from in situ analysis of Si isotopes, major and trace elements in silicates and metal.\ Geochimica et Cosmochimica Acta 267, 300.

\item[--] Krot, A.N., Connolly, H.C., Russel, S. \ 2018. Chondrules : records of protoplanetary disk processes. Cambridge University Press.
  
\item[--] Larimer, J.~W.\ 1975. The effect of C/O ratio on the condensation of planetary material. Geochim. Cosmochim. Acta 39,
389-392.

\item[--] Larimer, J.~W.\ 1979.\ The condensation and fractionation of refractory lithophile elements.\ Icarus 40, 446.

%\item[--] Libourel, G., Portail, M.,\ 2018.\ Overlooked Chondrules: A High Resolution Cathodoluminescence Survey.\ LPI Contributions 2067, 6088. 

\item[--] Lodders, K., Fegley, B.\ 1993.\ Lanthanide and actinide chemistry at high C/O ratios in the solar nebula.\ Earth and Planetary Science Letters 117, 125.

\item[--] Lodders, K.\ 2000.\ An Oxygen Isotope Mixing Model for the Accretion and Composition of Rocky Planets.\ Space Science Reviews 92, 341.

\item[--] McDonough, W.F.  and Sun, S.-s., 1995. The composition of the Earth. Chem Geol 120, 223-253.
  
%\item[--] Moynier, F., Yin, Q.-Z., Schauble, E. 2011. Isotopic evidence of Cr partitioning into Earth's core. Science 331 (6023), 1417. 

\item[--] Morbidelli, A., et al.,\ 2016.\ Fossilized condensation lines in the Solar System protoplanetary disk.\ Icarus 267, 368-376. 

\item[--] Mysen, B.~O., Kushiro, I.\ 1988.\ Condensation, evaporation, melting, and crystallization in the primitive solar nebula: experimental data in the system MgO-SiO-H$_{2}$ to 1.0x10$^{-9}$bar and 1870C with variable oxygen fugacity..\ American Mineralogist 73, 1.

\item[--] Nanne, J.~A.~M., Nimmo, F., Cuzzi, J.~N., Kleine, T.\ 2019.\ Origin of the non-carbonaceous-carbonaceous meteorite dichotomy.\ Earth and Planetary Science Letters 511, 44.

%\item[--] Norris, C.~A., Wood, B.~J.,\ 2017.\ Earth's volatile contents established by melting and vaporization.\ Nature 549, 507-510. 
  
\item[--] Ogihara, M., et al., \ 2018.\ Formation of the terrestrial planets in the solar system around 1 au via radial concentration of planetesimals.\ Astron. Astrophys. 612, L5. 

%\item[--] Pape, J., et al., 2018. Time and Duration of Chondrule Formation: Constraints from 26Al-26Mg Ages of Individual Chondrules, Geochim. Cosmochim. Acta  doi: https://doi.org/10.1016/j.gca.2018.10.017

%\item[--] Pringle, E.~A., et al., 2014. Silicon isotopes in 
%angrites and volatile loss in planetesimals. Proc. Natl. 
% Acad. Sci. USA 111, 17029-17032.

%\item[--] Ricolleau, A., Fei, Y., Corgne, A., Siebert, J., 
%Badro, J.\ 2011.\ Oxygen and silicon contents of Earth's core %from high pressure metal-silicate partitioning experiments.\ %Earth and Planetary Science Letters 310, 409.

\item[--] Palme, H. and O'Neill, H. St. C., 2014. Cosmochemical estimates of mantle composition. Treatise on Geophysics Vol. 3. Elsevier  pp. 1-39.

\item[--] Pringle, E.A., Moynier, F., Savage, P.S., Badro, J. and Barrat, J.A., 2014. Silicon isotopes in angrites and volatile loss in planetesimals. PNAS, 48, 17029-17032

\item[--] Rubie, D.C., Nimmo, F., Melosh, H.J. 2015. Formation of the Earth's Core. In: Treatise on Geophysics Vol. 9 Elsevier  pp. 43-79.

\item[--] Sugiura, N., Fujiya, W.\ 2014.\ Correlated accretion ages and {\ensuremath{\in}}$^{54}$Cr of meteorite parent bodies and the evolution of the solar nebula.\ Meteoritics and Planetary Science 49, 772.
  
%\item[--] Suzuki, T.~K., et al., \ 2016.\ Evolution of 
%protoplanetary discs with magnetically driven disc winds.\ 
%Astron. Astroph. 596, A74. 

%\item[--] Villeneuve, J., et al., \ 2009.\ Homogeneous Distribution of $^{26}$Al in the Solar System from the Mg Isotopic Composition of Chondrules.\ Science 325, 985. 

\item[--] Yang, C.-C., et al.,\ 2017.\ Concentrating small particles in protoplanetary disks through the streaming instability.\ Astron. Astroph. 606, A80. 
  
\item[--] Yoneda, S., Grossman, L.,\ 1995.\ Condensation of CaO sbnd MgO sbnd Al $_{2}$O $_{3}$sbnd SiO $_{2}$ liquids from cosmic gases.\ Geochim. Cosmochim. Acta 59, 3413-3444. 

\item[--] Youdin, A.~N., Goodman, J.,\ 2005.\ Streaming Instabilities in Protoplanetary Disks.\ Astron. J. 620, 459-469. 

\item[--] Young, E., et al., J., 2019. Near-equilibrium isotope fractionation during planetesimal evaporation. Icarus, 323, 1.

%\item[--] Wang, H.~S., Lineweaver, C.~H., Ireland, T.~R.\ 2018.\ The elemental abundances (with uncertainties) of the most Earth-like planet.\ Icarus 299, 460.

\end{itemize}



\section*{Supplementary material}

Table S1: Data used in constructing Fig. 1

| wt. % | Na | Mg | Al | Si | S | Cr | Fe | |
|---|---|---|---|---|---|---|---|---|
| CI | 0.496 | 9.54 | 0.84 | 10.7 | 5.9 | 0.262 | 18.66 | (1) |
| CM | 0.412 | 11.66 | 1.15 | 13.01 | 3.3 | 0.311 | 20.57 | (2) |
| CO | 0.415 | 13.77 | 1.37 | 15.41 | 2.0 | 0.338 | 23.98 | (2) |
| CV | 0.326 | 14.72 | 1.62 | 15.8 | 2.2 | 0.347 | 23.23 | (2) |
| R | 0.648 | 12.81 | 1.08 | 16.97 | 4.07 | 0.357 | 24.2 | (4) |
| H | 0.64 | 14.0 | 1.13 | 16.9 | 2.0 | 0.366 | 27.5 | (3) |
| L | 0.70 | 14.9 | 1.22 | 18.5 | 2.2 | 0.388 | 21.5 | (3) |
| LL | 0.70 | 15.3 | 1.19 | 18.9 | 2.3 | 0.374 | 18.5 | (3) |
| EH | 0.68 | 10.6 | 0.81 | 16.7 | 5.8 | 0.315 | 29.0 | (3) |
| EL | 0.58 | 14.1 | 1.05 | 18.6 | 3.3 | 0.305 | 22.0 | (3) |
| Bulk Earth no Si in core | 0.18 | 15.08 | 1.62 | 14.43 | 0.56 | 0.41 | 31.48 | (1) |
| 7%Si in core | 0.18 | 15.08 | 1.62 | 16.67 | 0.56 | 0.41 | 31.48 | (1) |

 (1) Palme and O'Neill (2014), assumed core fraction 32% with 85% Fe; (2) Wolf and Palme (2001), Na from Kallemeyn and Wasson (1981); (3) Wasson and Kallemeyn (1988) ; (4) Bischoff et al. 2011). All S data (except R-chondrites) are from Wasson and Kallemeyn (1981). Cr is calculated from a CI-bulk Earth Mg/Cr ratio.


Additional literature:

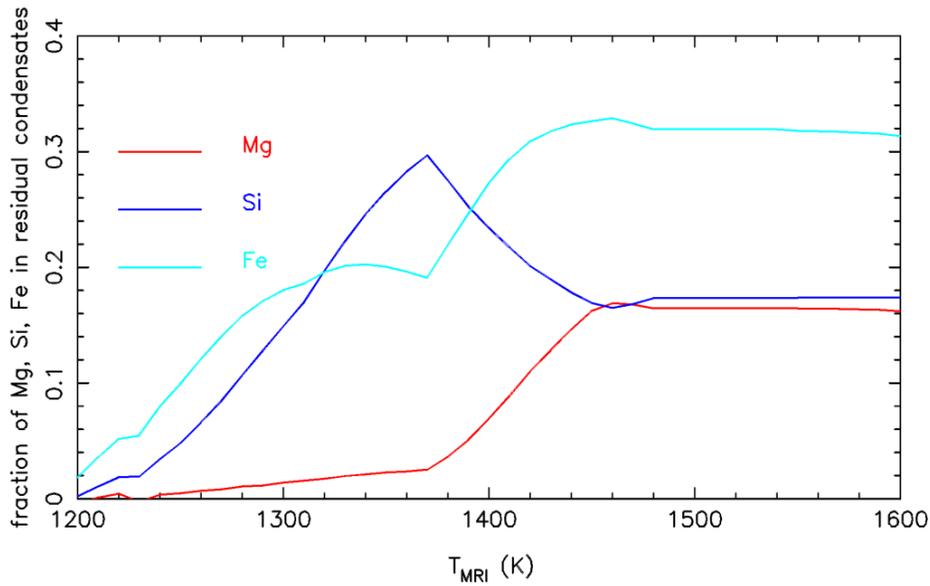

**Fig. S1**. The mass fractions $f^{res}_{Mg}$, $f^{res}_{Si}$ and $f^{res}_{Fe}$ of Mg, Si and Fe in residual condensates as a function of temperature of formation of the first planetesimals $T_{MRI}$. The fraction $f^{res}_{Al}$ is virtually 0 in the showed temperature range, so it is not plotted. These fractions are used in eq. (1) of the main paper.

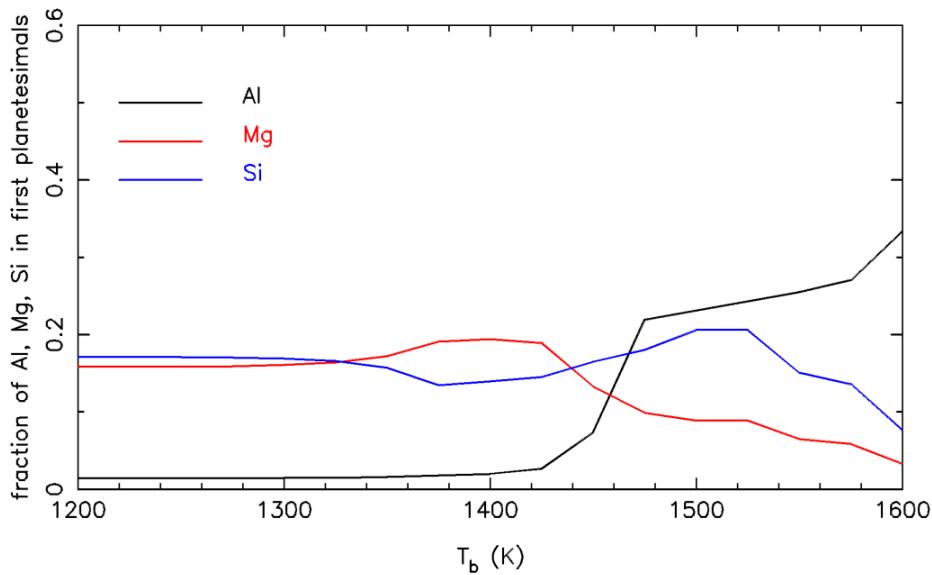

**Fig. S2**. The mass fractions $f^{first}_{Al}$, $f^{first}_{Mg}$ and $f^{first}_{Si}$ of Al, Mg and Si in first planetesimals as a function of the temperature of formation $T_b$. The quantities $f^{first}_{Al}$, $f^{first}_{Mg}$ and $f^{first}_{Si}$ are used in the modified eq. (1) in section 4.2 of the main paper.

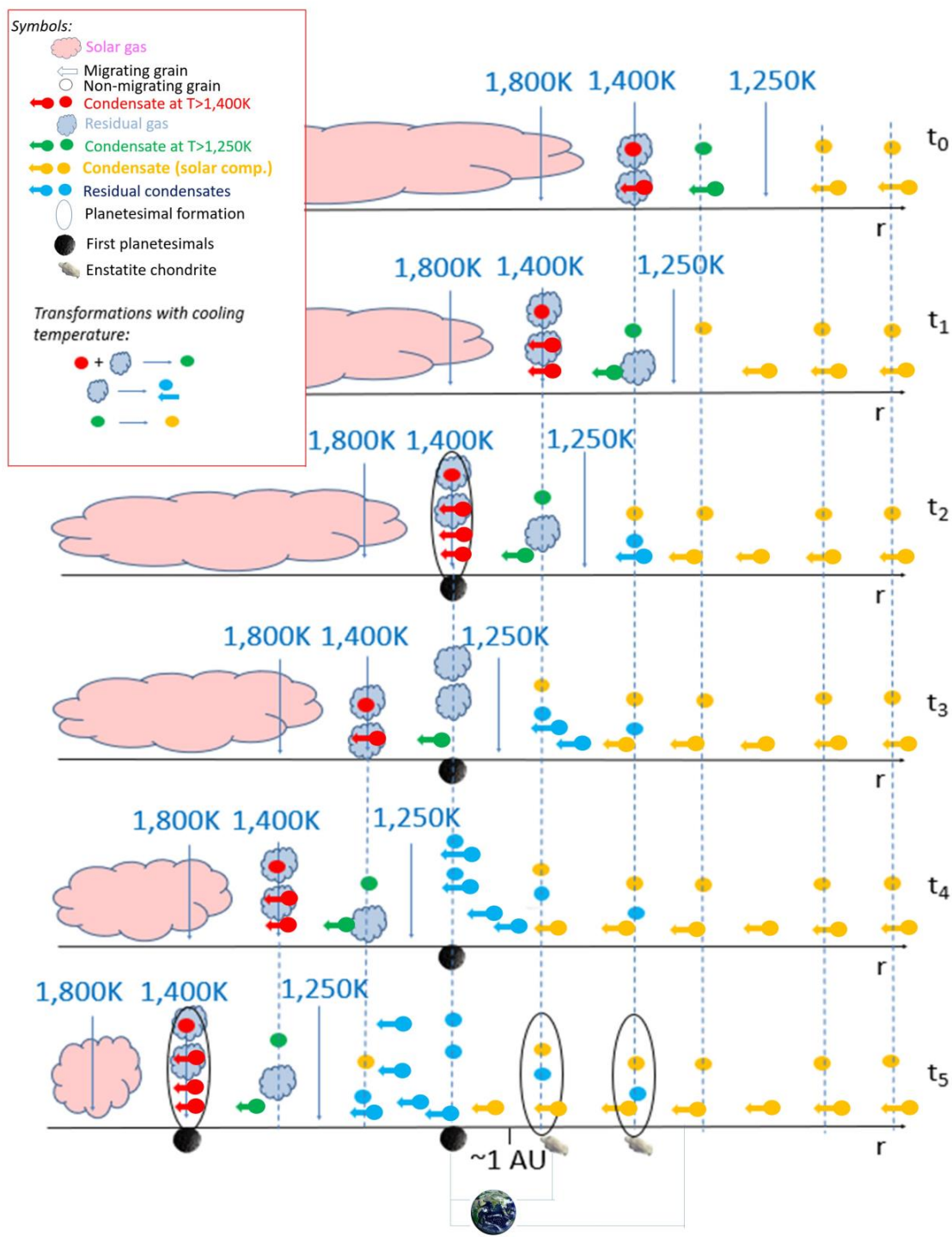

**Fig. S3:** Sketch of the global evolution of the disk envisioned in this work. The figure shows six time snapshots, from early ($t_0$) to late ($t_5$) going from top to bottom. The horizontal axis denotes distance from the star. The three vertical arrows labelled 1,800K, 1,400K, and 1,250K show the locations where the disk temperature has these values. As time progresses from one snapshot to the subsequent one, the disk cools, so the locations of these temperatures shift towards the sun (to the left).

At a temperature greater than 1,800K all material is in a gaseous form and the gas has a solar composition. At lower temperatures some materials have condensed to solids. We indicate in red the grains which followed the equilibrium condensation sequence down to a temperature T=1,400K, in green those which reached temperature T between 1,250K and 1,400K and in orange those which reached temperature T<1,250K. Orange grains have Al/Si and Mg/Si ratios as in CI-chondrites (solar values), although they may miss the volatile component (e.g. water) of CI meteorites if their condensation temperature is higher than a few hundred degrees.

Because we assume to be in a condensation sequence regime, the gas has to drift towards the star more slowly than the temperature lines, so that the parcels of gas cool over time. Thus, for simplification, we assume that the gas is at rest.

Grains drift towards the sun (leftwards) at different speeds, depending on their size[1]. For every condensation event we introduce two pairs of grains. One grain is assumed not to migrate and is represented by a filled dot. The other one instead migrates at the same speed as the temperature lines (filled dot with a left-pointing arrow). This is a simplification, of course, but it captures the real difference existing between grains that migrate faster than (or as fast as) the temperature lines and those migrating less fast. A guideline for reading the figure could be as follows. Consider first the red dot in the first snapshot ($t_0$). It has condensed at 1,400K. But because it does not move, it remains in equilibrium with the residual gas. As the temperature decreases ($t_1$) its composition changes to that of a grain condensed at 1,400K>T>1,250K (green dot on the $t_1$ panel located vertically below the red dot on the $t_0$ panel). Then, when the temperature declines further ($t_2$) it becomes a grain with a solar composition (orange dot). Similarly, the non-migrating green grain in the $t_0$ panel becomes a solar-composition orange dot in the $t_1$ panel (placed along the same vertical dashed line), after having been swept by the1,250K line. Now focus on migrating grains, represented by filled dots with arrows. First, consider the red one on the $t_0$ panel. It migrates with the 1,400K line, leaving behind the residual gas (represented by the gray cloud in the $t_1$ panel on the vertical dashed line passing through the 1,400K location on the $t_0$ panel). When the residual gas is swept by the 1,250K line ($t_2$ panel) it will condense into residual grains (pair of blue dots, one with an arrow, placed vertically along the dashed line). Similarly, the production of residual grains proceeds as more parcels of residual gas cool over time. Finally, look at what happens on the 1,400K line as time progresses. The migrating grain condensed at $t_0$ (red dot with arrow) remains on the temperature line (remember that this is not due to the specific assumption that the grain migrates at the same speed of the temperature line). Any grain migrating *faster* would remain trapped at the 1,400K line because it corresponds to the transition from a MRI disk to a quiescent disk, hence to a dust trap). So, at time $t_1$, on the 1,400K line there will be three grains: the migrating one from time $t_0$, and two new ones, formed at time $t_1$ because new solar gas is swept by the condensation line. Assuming again that half of the newly condensed grains migrate with the line, at time $t_2$ there will be 4 grains. This illustrates the pile-up process that eventually can lead to a solid/gas ratio large enough to trigger the streaming instability. Refractory-rich planetesimals can form this way. The figure shows two such events: one at $t_2$, the other at $t_5$ in order to illustrate that this process is cyclic and continues as long as the condition holds that the temperature lines migrate faster than the gas.

Late planetesimal formation can also occur when gas is depleted (possibly by photo-evaporation) so that the solid/gas ratio increases due to whatever grains remain in the disk. This is illustrated at time $t_5$, with the formation of enstatite chondrites from residual (blue) and solar-composition (orange) grains.

In this scenario, as explained in Sect. 4 of the main paper, the Earth builds from a combination of first, refractory-rich planetesimals, solar-composition grains and/or enstatite chondrites, with an excess of refractory material over solar-composition material.

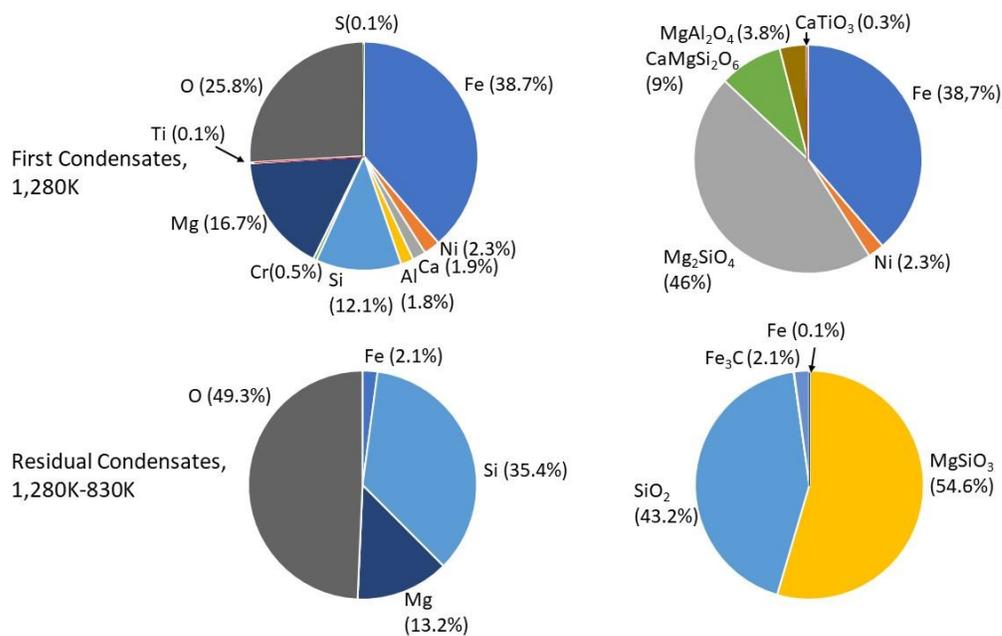

Fig. S4. The same as Fig. 6 of the main text, but computed for C/O=0.9 and $T_{MRI}$=1,280K

Table S2. Gas species included in the equilibrium condensation calculation from the factsage data base. http://www.crct.polymtl.ca/fact/documentation/

H(g)
H2(g)
C(g)
C2(g)
C3(g)
C4(g)
C5(g)
CH(g)
CH2(g)
CH3(g)
CH4(g)
C2H(g)
C2H2(g)
C2H3(g)
C2H4(g)
C2H5(g)
C2H6(g)
O(g)
O2(g)
O3(g)
OH(g)
H2O(g)
HOO(g)
HOOH(g)
CO(g)
C2O(g)
CO2(g)
C3O2(g)
HCO(g)
H2CO(g)
CH3O(g)
CH3O(g2)
CH3OH(g)
CH2CO(g)
C2H4O(g)
C2H4O(g2)
CH3CH2OH(g)
CH3CH2OH(g2)
COOH(g)
HCOOH(g)
CH3COOH(g)
Na(g)
Na2(g)
NaH(g)
NaO(g)
NaOH(g)
(NaOH)2(g)
Mg(g)
Mg2(g)
MgH(g)
MgO(g)
MgOH(g)
Mg(OH)2(g)
Al(g)
Al2(g)
AlH(g)
AlC(g)

AlC2(g)
Al2C2(g)
AlO(g)
AlO2(g)
Al2O(g)
Al2O2(g)
Al2O3(g)
AlOH(g)
AlOH(g2)
OAlOH(g)
Si(g)
Si2(g)
Si3(g)
SiH(g)
SiH4(g)
Si2H6(g)
SiC(g)
SiC2(g)
Si2C(g)
SiO(g)
SiO2(g)
S(g)
S2(g)
S3(g)
S4(g)
S5(g)
S6(g)
S7(g)
S8(g)
HS(g)
H2S(g)
H2S2(g)
CS(g)
CS2(g)
CH3SH(g)
C2H4S(g)
(CH3)2S(g)
(CH3)2S(g2)
CH3SSCH3(g)
SO(g)
SO2(g)
SO3(g)
SSO(g)
H2SO4(g)
COS(g)
C2H4OS(g)
(CH3)2SO(g)
(CH3)2SO2(g)
Na2SO4(g)
MgS(g)
AlS(g)
Al2S(g)
Al2S2(g)
SiS(g)
SiS2(g)
Ca(g)
Ca2(g)

CaH(g)
CaO(g)
CaOH(g)
Ca(OH)2(g)
CaS(g)
Ti(g)
TiO(g)
TiO2(g)
TiS(g)
Cr(g)
CrO(g)
CrO2(g)
CrO3(g)
CrOH(g)
CrOOH(g)
Cr(OH)2(g)
CrO2OH(g)
CrO(OH)2(g)
Cr(OH)3(g)
CrO2(OH)2(g)
CrO(OH)3(g)
Cr(OH)4(g)
CrO(OH)4(g)
Cr(OH)5(g)
Cr(OH)6(g)
CrS(g)
Fe(g)
FeO(g)
Fe(OH)2(g)
Fe(CO)5(g)
FeS(g)
Ni(g)
NiH(g)
NiO(g)
Ni(OH)2(g)
Ni(CO)4(g)
NiS(g)

Table S3. Pure solid phases included in the equilibrium condensation calculation from the factsage data base. http://www.crct.polymtl.ca/fact/documentation/

C(s)
C(s2)
H2O(s)
Na(s)
NaH(s)
C2Na2(s)
NaO2(s)
Na2O2(s)
Na2O2(s2)
NaOH(s)
NaOH(s2)
Na2CO3(s)
Na2CO3(s2)
Na2CO3(s3)
NaHCO3(s)
Na2CO3(H2O)10(s)
Mg(s)
MgH2(s)
MgC2(s)
Mg2C3(s)
Mg(OH)2(s)
MgCO3(s)
Al(s)
AlH3(s)
Al4C3(s)
Al(OH)3(s)
Al2O3(H2O)(s)
Al2O3(H2O)(s2)
Al2O3(H2O)3(s)
AlCH2NaO5(s)
Si(s)
SiC(s)
SiC(s2)
H2SiO3(s)
H4SiO4(s)
H2Si2O5(s)
H6Si2O7(s)
Mg2Si(s)
Mg3Si2O5(OH)4(s)
Mg3Si4O10(OH)2(s)
Mg7Si8O22(OH)2(s)
Al2SiO5(s)
Al2SiO5(s2)
Al2SiO5(s3)
(Al2O3)(SiO2)2(H2O)2(s)
(Al2O3)(SiO2)2(H2O)2(s2)
Al2Si4O10(OH)2(s)
NaAlSi2O6H2O(s)
NaAl3Si3O12H2(s)
Mg5Al2Si3O10(OH)8(s)
Mg2Al4Si5O18(H2O)(s)
Na2Mg3Al2Si8O22(OH)2(s)

SO3(s)
Na2S(s)
NaS2(s)
Na2S2(s)
Na2S3(s)
Na2SO3(s)
Na2SO4(s)
Na2SO4(s2)
Na2SO4(s3)
Na2SO4(H2O)10(s)
Na3(OH)(SO4)(s)
MgS(s)
MgSO4(s)
MgSO4(H2O)(s)
MgSO4(H2O)6(s)
MgSO4(H2O)7(s)
Al2S3(s)
Al2(SO4)3(s)
Al2(SO4)3(H2O)6(s)
SiS(s)
SiS2(s)
Ca(s)
Ca(s2)
CaH2(s)
CaH2(s2)
CaC2(s)
CaC2(s2)
CaO2(s)
Ca(OH)2(s)
CaCO3(s)
CaCO3(s2)
CaC2O4(H2O)(s)
Mg2Ca(s)
CaMg(CO3)2(s)
Al2Ca(s)
Al4Ca(s)
(CaO)3(Al2O3)(H2O)6(s)
(CaO)4(Al2O3)(H2O)13(s)
CaSi(s)
CaSi2(s)
Ca2Si(s)
(CaO)(SiO2)2(H2O)2(s)
(CaO)3(SiO2)2(H2O)3(s)
(CaO)4(SiO2)6(H2O)5(s)
(CaO)5(SiO2)6(H2O)3(s)
(CaO)6(SiO2)6(H2O)(s)
(CaO)8(SiO2)6(H2O)3(s)

(CaO)10(SiO2)12(H2O)11(s)
(CaO)10(SiO2)12(H2O)21(s)
(CaO)12(SiO2)6(H2O)7(s)
Ca2Mg5Si8O22(OH)2(s)
CaAl2Si2O7(OH)2(H2O)(s)
CaAl4Si2O10(OH)2(s)
CaAl14Si22O60(OH)12(s)
Ca2Al3Si3O12(OH)(s)
Ca2Al3Si3O12(OH)(s2)
(CaO)2(Al2O3)2(SiO2)8(H2O)7(s)
CaS(s)
CaSO3(s)
CaSO4(s)
CaSO4(s2)
CaSO3(H2O)2(s)
CaSO4(H2O)2(s)
(CaSO3)2(H2O)(s)
(CaSO4)2(H2O)(s)
Ti(s)
Ti(s2)
TiH2(s)
CTi(s)
TiO(s)
TiO(s2)
TiAl(s)
TiAl3(s)
SiTi(s)
Si2Ti(s)
TiS(s)
TiS2(s)
TiS3(s)
Ti2S(s)
Ti2S3(s)
Cr(s)
Cr3C2(s)
Cr4C(s)
Cr7C3(s)
Cr23C6(s)
CrO2(s)
CrO3(s)
Cr5O12(s)
Cr8O21(s)
Cr(CO)6(s)
CrSi2(s)
Cr3Si(s)
Cr5Si3(s)
Cr2(SO4)3(s)
Fe(s)
Fe(s2)

Fe3C(s)
Fe3C(s2)
Fe(OH)2(s)
Fe(OH)3(s)
Fe2O3(H2O)(s)
FeCO3(s)
FeAl3(s)
FeSi(s)
FeSi2(s)
Fe3Si(s)
Fe3Si7(s)
Fe7Si8O22(OH)2(s)
FeSO4(s)
Fe2(SO4)3(s)
FeSO4(H2O)7(s)
FeTi(s)
Fe2Ti(s)
Ni(s)
Ni2H(s)
Ni3C(s)
NiOOH(s)
Ni(OH)2(s)
NiO2(H2O)(s)
NiCO3(s)
MgNi2(s)
NiAl(s)
Al3Ni(s)
Ni2Al3(s)
Ni3Al(s)
NiSi(s)
Ni2Si(s)
Ni7Si13(s)
NiSO4(s)
NiSO4(H2O)(s)
NiSO4(H2O)6(s)
NiSO4(H2O)7(s)
Ni4SO4(OH)6(s)
NiTi(s)
NiTi2(s)
Ni3Ti(s)
(NiO)(TiO2)(s)
NiTi2O5(s)
Ni2TiO4(s)
S(s)
S(s2)
Cr2S3(s)
FeS(s)
FeS2(s)
Fe7S8(s)
Fe9S10(s)
Fe10S11(s)
Fe11S12(s)
FeCr2S4(s)
NiS(s)
NiS2(s)
Ni3S2(s)
Ni3S4(s)

Ni7S6(s)
Ni9S8(s)
MgO(s)
Al2O3(s)
Al2O3(s2)
Al2O3(s3)
Al2O3(s4)
SiO2(s)
SiO2(s2)
SiO2(s3)
SiO2(s4)
SiO2(s5)
SiO2(s6)
SiO2(s7)
SiO2(s8)
MgSiO3(s)
MgSiO3(s2)
MgSiO3(s3)
MgSiO3(s4)
MgSiO3(s5)
MgSiO3(s6)
MgSiO3(s7)
Mg2SiO4(s)
Mg2SiO4(s2)
Mg2SiO4(s3)
Al2Si2O7(s)
NaAlSiO4(s)
NaAlSiO4(s2)
NaAlSi2O6(s)

NaAlSi3O8(s)
NaAlSi3O8(s2)
Mg4Al10Si2O23(s)
Mg3Al2Si3O12(s)
Mg2Al4Si5O18(s)
CaO(s)
CaAl2O4(s)
CaAl4O7(s)
CaAl12O19(s)
Ca3Al2O6(s)
CaMg2Al16O27(s)
Ca2Mg2Al28O46(s)
Ca3MgAl4O10(s)
CaSiO3(s)
CaSiO3(s2)
Ca2SiO4(s)
Ca2SiO4(s2)
Ca2SiO4(s3)
Ca3SiO5(s)
Ca3Si2O7(s)
CaMgSi2O6(s)
Ca2MgSi2O7(s)
Ca3MgSi2O8(s)
CaAl2SiO6(s)
CaAl2Si2O8(s)
CaAl2Si2O8(s2)
Ca2Al2SiO7(s)
Ca3Al2Si3O12(s)
TiO2(s)

TiO2(s2)
Ti2O3(s)
Ti2O3(s2)
Ti3O5(s)
Ti3O5(s2)
Ti4O7(s)
Ti5O9(s)
Ti6O11(s)
Ti7O13(s)
Ti8O15(s)
Ti9O17(s)
Ti10O19(s)
Ti20O39(s)
MgTiO3(s)
Mg2TiO4(s)
Mg2TiO4(s2)
MgTi2O5(s)
Al2TiO5(s)
Al4TiO8(s)
CaTiO3(s)
CaTiO3(s2)
Ca2Ti2O5(s)
Ca2Ti2O5(s2)
Ca3Ti2O6(s)
Ca3Ti2O7(s)
Ca5Ti4O13(s)
CaSiTiO5(s)
Cr2O3(s)
CaCr2O4(s)

CaCr2O4(s2)
(Ca2Cr3)Cr10O20(s)
(CaCr)Si4O10(s)
Ca3Cr2Si3O12(s)
Fe2O3(s)
Fe2O3(s2)
Fe2O3(s3)
Al2Fe2O6(s)
FeSiO3(s)
FeSiO3(s2)
FeSiO3(s3)
Fe2SiO4(s)
Fe2SiO4(s2)
Fe2SiO4(s3)
Fe2Al4Si5O18(s)
Fe3Al2Si3O12(s)
CaFe2O4(s)
Ca2Fe2O5(s)
CaFe4O7(s)
CaFeSi2O6(s)
Ca2FeSi2O7(s)
Ca3Fe2Si3O12(s)
(FeO)(TiO2)(s)
FeTi2O4(s)
FeTi2O5(s)
(FeO)2(TiO2)(s)
NiO(s)
Ni2SiO4(s)
NiCaSi2O6(s)

*Note: (s2, s3, ...) standing for different polymorphs of the (s) phase*


\end{document}